\documentclass[fleqn,usenatbib,onecolumn]{mnras}
\usepackage[T1]{fontenc}
\usepackage{amsmath,amssymb}
\usepackage{graphicx}
\usepackage{subfloat}
\usepackage{subfig}

\DeclareRobustCommand{\VAN}[3]{#2}
\let\VANthebibliography\thebibliography
\def\thebibliography{\DeclareRobustCommand{\VAN}[3]{##3}\VANthebibliography}
\pubyear{2024}

\begin{document}

\title[Quantum refraction in pulsar emission]{Quantum refraction effects in
pulsar emission}
\author[D. H. Kim et al.]{ Dong-Hoon Kim,$^{1,2}$\thanks{%
E-mail: ki13130@gmail.com} Chul Min Kim$^{3}$ and Sang Pyo Kim$^{4,5}$ \\
$^{1}$The Research Institute of Basic Science, Seoul National University,
Seoul 08826, Republic of Korea\\
$^{2}$Department of Physics and Astronomy, Seoul National University, Seoul
08826, Republic of Korea\\
$^{3}$Advanced Photonics Research Institute, Gwangju Institute of Science
and Technology, Gwangju 61005, Republic of Korea\\
$^{4}$Department of Physics, Kunsan National University, Gunsan 54150,
Republic of Korea\\
$^{5}$Asia Pacific Center for Theoretical Physics, Pohang 37673, Republic of
Korea }
\date{Accepted XXX. Received YYY; in original form ZZZ}
\maketitle

\begin{abstract}
Highly magnetized neutron stars exhibit the vacuum non-linear electrodynamics effects, 
which can be well described using the one-loop effective action for quantum electrodynamics. In this context, 
we study the propagation and polarization of pulsar radiation, based on the post-Maxwellian Lagrangian from the 
Heisenberg-Euler-Schwinger action. Given the refractive index obtained 
from this Lagrangian, we determine the leading-order corrections to both the 
propagation and polarization vectors due to quantum refraction via perturbation 
analysis. In addition, the effects on the orthogonality between the propagation and 
polarization vectors and the Faraday rotation angle, all due to quantum refraction 
are investigated. Furthermore, from the dual refractive index and the associated polarization modes, 
we discuss quantum birefringence, with the optical phenomenology analogous to its classical counterpart.
\end{abstract}

\label{firstpage} \pagerange{\pageref{firstpage}--\pageref{lastpage}}





\section{Introduction}

\label{introduction}

Neutron stars have strong magnetic fields on their surface from $10^8 
\,\mathrm{G}$ up to $10^{15} \,\mathrm{G}$, and in particular magnetars have the
strongest magnetic fields in the universe with $10^{13} - 10^{15} 
\,\mathrm{G}$, which are near or a little above the supercritical value $B_\mathrm{c} = m_\mathrm{e}^2 c^3/e\hbar = 4.4
\times 10^{13} \,\mathrm{G}$ (\cite{Olausen2014McGILL,Kaspi2017Magnetars}). In
such strong magnetic fields, the vacuum becomes a polarized medium due to the interaction of the fields with virtual electron-positron pairs. 
As a consequence, a photon propagating in the strong magnetic field background can be refracted or split, which is prohibited in the classical Maxwell theory. 

The critical magnetic field (or the so-called Schwinger field) is three order
higher than the current highest strength achieved with ultra-intense lasers; i.e., 
$B = 4.9 \times 10^{-4} B_\mathrm{c}$ (\cite{Yoon2021Realization}). Therefore, highly
magnetized neutron stars will provide a celestial laboratory to test quantum electrodynamics (QED) in the strong field regime and the relevant consequences 
(for review and references, see \cite{Ruffini2010Electronpositron,Fedotov:2022ely,Hattori:2023egw}). Recently,
the surface magnetic field for Swift J0243.6+6124 has been directly measured
from the detection of cyclotron resonance scattering (\cite{2022ApJ...933L...3K}%
). Also, space missions have been proposed to investigate the strong-field QED effects: 
the enhanced X-ray Timing and Polarimetry (eXTP) (\cite{Santangelo2019Physics}) and the Compton Telescope project (\cite%
{Wadiasingh2019Magnetars}).

In this paper, we study the propagation and polarization of a photon in the dipole magnetic field
background of a pulsar model, based on the post-Maxwellian (PM) Lagrangian; it is, in the weak field limit, 
the generic form of non-Maxwellian Lagrangian for the non-linear vacuum (\cite{Sorokin2022Introductory}) and 
has been used to test vacuum polarization effects in the PVLAS (Polarizzazione del Vuoto con LAser) project (\cite{Ejlli2020PVLASa}). 
The Heisenberg-Euler-Schwinger (HES) Lagrangian of strong-field QED (\cite{Heisenberg1936Folgerungen,Schwinger1951Gauge}) is also well approximated by the PM Lagrangian below the critical field strength. 
Here, we develop the recent analysis of vacuum birefringence in \cite{Kim2023Vacuum} further for a practical model of pulsar emission, 
where the magnetic field is defined for an oblique rotator with an inclination angle and the electric field is discarded. We consider the dual refractive index and 
the associated polarization vectors of a probe photon (\cite{Kim202231}) to investigate the photon propagation in the strong
magnetic field background of this pulsar model. Then the leading-order corrections to both the propagation and polarization vectors due to quantum refraction are determined via perturbation analysis. 
Our study provides a novel, complementary approach to and an elaboration of other similar studies in \cite%
{Heyl2000Polarization,Heyl2002QED,Heyl2003highenergy,Caiazzo2018Vacuum,Heyl2018Strongly,Caiazzo2019QED,Caiazzo2021Polarization,Caiazzo2022Probing}.

The paper is organized as follows. In Section \ref{qvn} we review the Lagrangian formalism for the non-linearity of the quantum vacuum due to strong electromagnetic fields; 
a brief account of the HES Lagrangian and the PM Lagrangian as its approximation is given. In Section \ref{dlr} we work out the deflection of a light ray from pulsar emission due to quantum refraction; the leading-order corrections to the propagation vector and then to the trajectory of the light ray are determined. 
In Section \ref{cplr} we look into the dual refractive index and the associated polarization modes of the light ray under the effect of quantum refraction; 
the leading-order corrections to the polarization vectors for Case I (Section \ref{casei}) and Case II (Section \ref{caseii}) are determined. In addition, 
the effects on the orthogonality between the propagation and polarization vectors and the Faraday rotation angle, due to quantum refraction
are investigated. Furthermore, in regard to the optical phenomenology from the dual refractive index and the associated polarization modes, we discuss quantum birefringence for this pulsar emission. 
Then finally, we conclude the paper with discussions on other similar studies and follow-up studies.

\section{Non-linear Electrodynamics due to Strong Fields}
\label{qvn}
Non-linear electrodynamic (NED) effects of the vacuum in the presence of strong electromagnetic fields have been studied employing the \textit{effective Lagrangian} formalism. 
One of the most well-known NED models is the Heisenberg-Euler-Schwinger (HES) Lagrangian, which is obtained by adding the one-loop QED correction to the Maxwell Lagrangian, due to spin-1/2 fermions of mass $m_\mathrm{e}$ and charge $e$ in electromagnetic fields of arbitrary strengths (\cite{Heisenberg1936Folgerungen,Schwinger1951Gauge}): in the convention with $\hbar =c=1$, 
\begin{equation}
\mathcal{L}_{\mathrm{HES}}\left( a,b\right) =\mathcal{L}^{\left(0\right)
}\left( a,b\right) +\mathcal{L}^{\left(1\right) }\left( a,b\right) =\frac{%
b^{2}-a^{2}}{2}+\mathcal{L}^{\left(1\right) }\left( a,b\right), \label{hes} 
\end{equation}%
where $\mathcal{L}^{\left(0\right)}\left( a,b\right)$ refers to the classical Maxwell Lagrangian, defined through
\begin{equation}
a\equiv \sqrt{\sqrt{F^{2}+G^{2}}+F},~b\equiv \sqrt{\sqrt{F^{2}+G^{2}}-F}, \label{ab}
\end{equation}%
with the Lorentz- and gauge-invariant Maxwell scalar $F$ and pseudo-scalar $G$
\begin{equation}
F\equiv \frac{1}{4}F^{\mu \nu }F_{\mu \nu }=\frac{1}{2}\left( \mathbf{B}^{2}-%
\mathbf{E}^{2}\right) =\frac{1}{2}\left( a^{2}-b^{2}\right),~G\equiv \frac{1%
}{4}F^{\mu \nu }F_{\mu \nu }^{\ast }=-\mathbf{E}\cdot \mathbf{B}=\mathrm{sign%
}\left( \pm \right) ab, \label{FG}
\end{equation}%
and $\mathcal{L}^{\left(1\right) }\left( a,b\right)$ refers to the Lagrangian of one-loop correction,
\begin{equation}
\mathcal{L}^{\left(1\right) }\left( a,b\right) =-\frac{1}{8\pi ^{2}}%
\int_{0}^{\infty }\mathrm{d}s\,\frac{e^{-m_\mathrm{e}^{2}s}}{s^{3}}\left\{ \left(
eas\right) \coth \left( eas\right) \left( ebs\right) \cot \left( ebs\right) -%
\left[ 1+\frac{\left( eas\right) ^{2}-\left( ebs\right) ^{2}}{3}\right]
\right\}. \label{hes1}
\end{equation}%

However, in the weak-field limit (below the critical field strength $B_\mathrm{c}$), the HES Lagrangian (\ref{hes}) has the leading-order contribution, 
the so-called post-Maxwellian (PM) Lagrangian (\cite{Euler1935Ueber}):\footnote{In effect, the PM Lagrangian approximates the HES Lagrangian well even up  to $B=0.14B_\mathrm{c}$, with an error less than 1\%. }
\begin{equation}
\mathcal{L}_{\mathrm{PM}}\left( a,b\right) =-F+\eta _{1}F^{2}+\eta _{2}G^{2}, \label{pm}
\end{equation}%
where $\eta _{1}$ and $\eta _{2}$ are parameters defined via $\eta_{1}/4=\eta _{2}/7=e^{4}/\left( 360\pi ^{2}m_\mathrm{e}^{4}\right) $.

 In some NED models, a parity violating term proportional to $F
 G$ is added to the PM Lagrangian (\cite{Ni2013Foundations}). However, for the rest of the paper, our analysis is based on the PM Lagrangian; the dual refractive index as given by equation (\ref{n1}) in Section \ref{dlr} is derived using this (\cite{Adler1971Photon,Kim202231}).

\section{Deflection of a light ray due to quantum refraction}

\label{dlr}

A light ray is defined as an orthogonal trajectory to the geometrical
wave-front $\mathcal{S}\left( x,y,z\right) =\mathrm{const.}$, and therefore
can be described by%
\begin{equation}
n\frac{\mathrm{d}\mathbf{r}}{\mathrm{d}s}=\nabla \mathcal{S},  \label{lr1}
\end{equation}%
where $s$ is an affine parameter to measure the length of the ray and $%
n=n\left( \mathbf{r}\right) $ is the refractive index given as a function of
the position $\mathbf{r}$ on the ray (\citet{Born1999Principles}). It can be
further shown that 
\begin{equation}
\frac{\mathrm{d}}{\mathrm{d}s}\left( n\frac{\mathrm{d}\mathbf{r}}{\mathrm{d}s%
}\right) =\nabla n.  \label{lr2}
\end{equation}%
Let $\mathbf{\hat{n}}\equiv \mathrm{d}\mathbf{r/}\mathrm{d}s$ be the unit
propagation vector for the light ray emitted from a spot either at rest or
in motion at a constant velocity.\footnote{%
This is in contrast with our actual case, wherein the emission spot itself
is under the centrifugal acceleration due to the rotation of a pulsar
magnetosphere, as described by equation (\ref{mhd}).} Then equation (\ref{lr2}) leads
to (\citet{Born1999Principles}) 
\begin{equation}
\mathbf{\hat{n}}=\left\{ 
\begin{array}{ll}
\mathrm{const.} & \text{for }n=\mathrm{const.}, \\ 
\frac{\int \nabla n\,\mathrm{d}s}{n} & \text{for }n\neq \mathrm{const.}.%
\end{array}%
\right.  \label{lrd}
\end{equation}%
The expression of $\mathbf{\hat{n}}$ for $n\neq \mathrm{const.}$ can be
applied to a mechanism of how the light ray is deflected, for example, due
to the quantum refraction effect in pulsar emission, as will be described below. In the
presence of the effect, the refractive index $n$ is given by a function of
the position $\mathbf{r}$, at which the light ray crosses a local magnetic
field line in a pulsar magnetosphere; otherwise, it would simply be a
constant.

\begin{figure*}
\centering
\includegraphics[width=\textwidth]{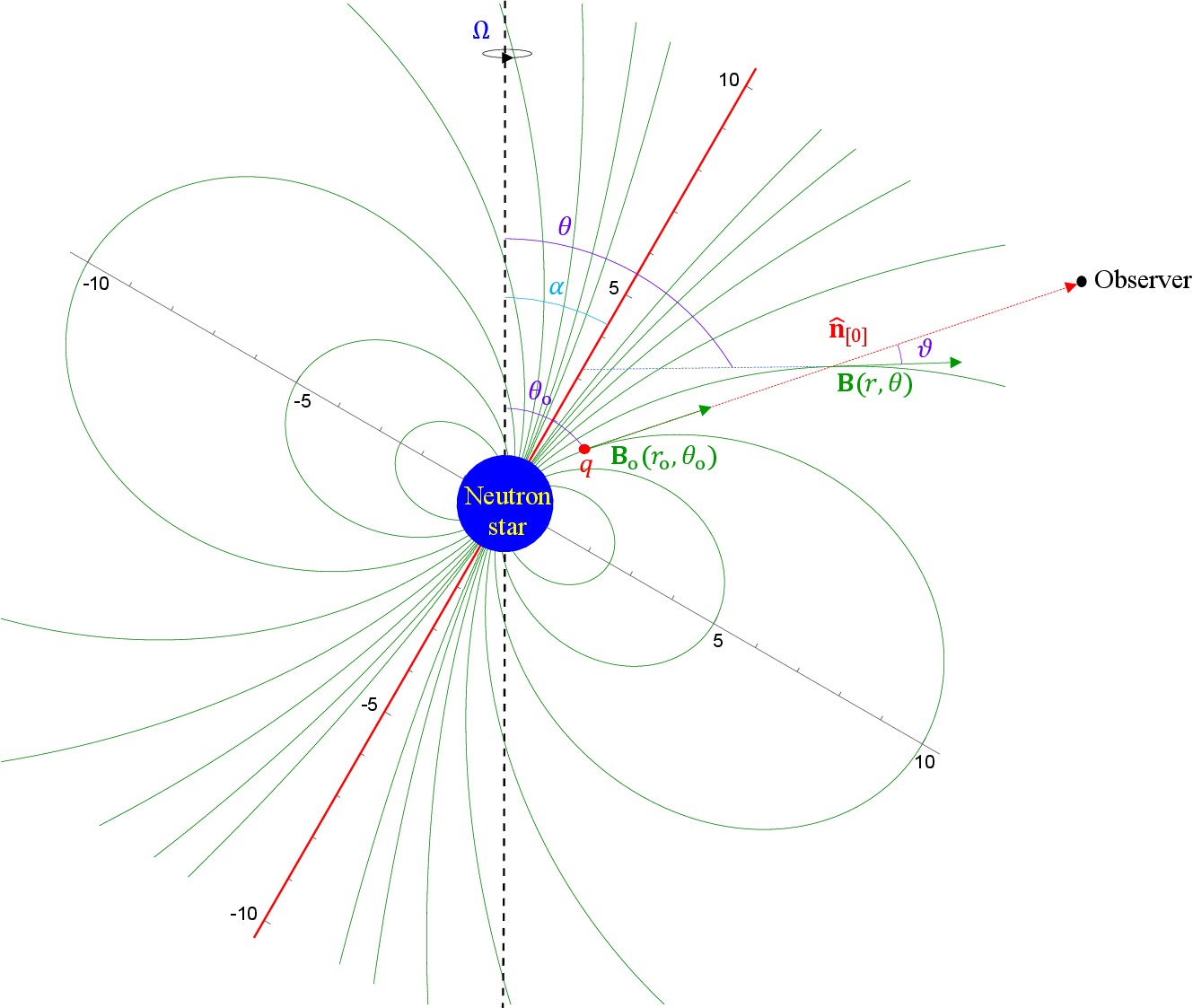}
\caption{A cross-sectional view of a pulsar magnetosphere with the dipole
magnetic field lines (green) around a neutron star. The vertical dashed line (black)
and the inclined solid line (red) represent the rotation axis and the
magnetic axis, respectively. $\alpha$ between these axes denotes
the inclination angle. The scale of the unity in this graph is
equivalent to the neutron star radius $\sim 10^{6}\,\mathrm{cm}$. The red
dotted line represents the trajectory curve of the light ray traced by $\mathbf{%
\hat{n}}_{\left[ 0\right] }$ as projected onto the $xz$-plane. (Credit: 
\citet{Kim2021General}, reproduced with modifications.) }
\label{fig1}
\end{figure*}

According to \cite{Kim202231}, the refractive index $n$ can be derived using the PM Lagrangian (\ref{pm}) as
\begin{equation}
n=\left\{ 
\begin{array}{ll}
\sqrt{\frac{1-\left(\eta _{1}-2\eta _{2}\right)B^{2}}{1-\left(\eta _{1}-2\eta
_{2}\cos ^{2}\vartheta\right)B^{2}}} & \text{for Case I}, \\ 
\sqrt{\frac{1-\eta _{1}B^{2}}{1-\left(\eta _{1}+2\eta _{1}\sin
^{2}\vartheta\right)B^{2}}} & \text{for Case II},%
\end{array}%
\right.  \label{n1}
\end{equation}%
where $B$ is the local magnetic field strength at a point in a pulsar magnetosphere, and $%
\vartheta $ denotes the angle between the light ray trajectory and the local
magnetic field line (see Fig. \ref{fig1}). Here we have named Case I and
Case II for two different values of the refractive index attributed to the
same point in the magnetosphere; the propagation and polarization of the
light ray are associated with these values. Later in Sections \ref{casei} and \ref{caseii}, the two polarization vectors for Case I and Case II, 
as given by (\ref{e0}) and (\ref{e00}), respectively are set to be orthogonal to each other 
and to the propagation vector such that the three vectors form a classical orthonormal basis. 
However, for the rest of this Section, we focus on Case I as there is little difference in the propagation of the
light ray between the two cases. Then in Section \ref{cplr} we look into the
polarization of the light ray for both the cases and discuss quantum
birefringence in relation to it.

One should note that the refractive index $n$ as given by (\ref{n1}) has no dependence on the frequency of radiation. 
Consequently, in our entire analysis, the quantum refraction effects derived from this, on the propagation and polarization of a photon 
have no frequency dependence either, as can be checked with equations (\ref{n3}) and (\ref{pol4}), respectively later. However, in order for 
equation (\ref{n1}) to be considered valid, the frequency of pulsar radiation must be significantly lower than that for excitation of 
the quantum vacuum ($\sim 10^{20}\,\mathrm{Hz}$; in the gamma-ray regime), which corresponds to the photon energy 
required to create an electron-positron pair, such that the vacuum is far from resonance. In addition, the plasma effects may be neglected 
if the pulsar radiation frequency is much higher than the local plasma frequency ($\sim 10^{9}\, \mathrm{Hz}$) (\cite{Petri2016Theory}). 
Therefore, our pulsar radiation can be safely assumed to cover optical to X-ray emissions ($\sim 10^{12}$ to $10^{17}\,\mathrm{Hz}$) in this work.  

One can expand the refractive index $n$ for Case I in equation (\ref{n1}), having 
$\eta _{1}$, $\eta _{2}\sim 10^{-31}\,\mathrm{g}^{-1}\,\mathrm{cm\,s}^{2}$
and $B<B_{\mathrm{c}}$ (critical magnetic field)$\,\sim 10^{13}\,\mathrm{G}$%
, and thus $\eta _{1}B^{2}$, $\eta _{2}B^{2}\ll 1$.\ Then it can be
approximated as (\cite{Adler1971Photon})
\begin{equation}
n\approx 1+\eta _{2}B^{2}\sin ^{2}\vartheta +\mathcal{O}\left( \eta
_{1}^{2}B^{4},\eta _{1}\eta _{2}B^{4},\eta _{2}^{2}B^{4}\right) ,  \label{n2}
\end{equation}%
which implies that the term $\eta _{2}B^{2}\sin ^{2}\vartheta \sim
10^{-4}\left( B/B_{\mathrm{c}}\right) ^{2}$ is the leading order quantum
correction to $n=1$ for classical optics, while $\mathcal{O}\left( \eta
_{1}^{2}B^{4},\eta _{1}\eta _{2}B^{4},\eta _{2}^{2}B^{4}\right) $ means the
next-to-leading order terms to be ignored in our analysis. For computational
purposes, the correction can be treated as the leading order perturbation
with $\eta _{2}B^{2}$ being a perturbation parameter. It should be noted
here that $n\rightarrow 1$, i.e., the refractive index goes back to the
classical limit as $B\rightarrow 0$ in the far field zone of the
magnetosphere.

Keeping equation (\ref{n2}) in mind, from equation (\ref{lrd}) one can determine the
deflection of the light ray to leading order via 
\begin{equation}
\delta \mathbf{\hat{n}}_{\left[ 1\right] }\approx \int \nabla \delta n_{ %
\left[ 1\right] }\,\mathrm{d}s+\mathcal{O}\left( _{\left[ 2\right] }\right) ,
\label{n3}
\end{equation}%
where $\delta \left( \cdots \right) _{\left[ 1\right] }$ means the leading
order quantum correction to the quantity $\left( \cdots \right) $, led by $%
\eta _{1}B^{2}$ or $\eta _{2}B^{2}$, and $\mathcal{O}\left( _{\left[ 2\right]
}\right) $ is a shorthand expression for $\mathcal{O}\left( \eta
_{1}^{2}B^{4},\eta _{1}\eta _{2}B^{4},\eta _{2}^{2}B^{4}\right) $. On the
other hand, by $\mathbf{\hat{n}}_{\left[ 0\right] }$ we mean the \textit{\
unperturbed }(classical) propagation direction of the light ray. A classical
trajectory of the light ray traced by $\mathbf{\hat{n}}_{\left[ 0\right] }$
is as represented by the red dashed line as in Fig. \ref{fig1}, wherein it
appears to be a straight line, being projected onto the $xz$-plane.

In our pulsar emission model, a light ray of curvature radiation is emitted from
the dipole magnetic field:%
\begin{equation}
\mathbf{B}\left( r,\theta ,\phi \right) =\frac{2\mu \left( \cos \alpha \cos
\theta +\sin \alpha \sin \theta \cos \phi \right) }{r^{3}}\mathbf{e}_{\hat{r}%
}+\frac{\mu \left( \cos \alpha \sin \theta -\sin \alpha \cos \theta \cos
\phi \right) }{r^{3}}\mathbf{e}_{\hat{\theta}}+\frac{\mu \sin \alpha \sin
\phi \,}{r^{3}}\mathbf{e}_{\hat{\phi}},  \label{B}
\end{equation}%
where $\mu$ is the magnetic dipole moment and $\alpha$ denotes the inclination angle 
between the rotation axis and the magnetic axis, and the light ray is tangent to the field line 
at the emission point $\left( x_{\mathrm{o}},y_{%
\mathrm{o}},z_{\mathrm{o}}\right) =\left( r_{\mathrm{o}}\sin \theta _{%
\mathrm{o}},0,r_{\mathrm{o}}\cos \theta _{\mathrm{o}}\right)$ 
(see Fig. \ref%
{fig1}). At the same time, however, our pulsar magnetosphere rotates, and
the magnetic field lines get twisted due to the magneto-centrifugal
acceleration on the plasma particles moving along the field lines (%
\citet{Blandford1982Hydromagnetic}). Taking into consideration this
magnetohydrodynamic (MHD) effect due to rotation, one can describe
classically the propagation direction of the light ray, which must line up
with the particle velocity in order for an observer to receive the
radiation, as (\citet{Gangadhara2005method}) 
\begin{equation}
\mathbf{\hat{n}}_{\left[ 0\right] }=\beta \mathbf{\hat{B}}+\frac{\mathbf{\Omega }%
\times \mathbf{r}}{c},  \label{mhd}
\end{equation}%
where on the right-hand side $\mathbf{\hat{B}}\equiv\mathbf{B/}\left\vert \mathbf{B}\right\vert$ 
and the second term accounts for the centrifugal acceleration, with 
$\mathbf{\Omega }\equiv \Omega \mathbf{e}_{z}$ and $%
\Omega $ being a pulsar rotation frequency, and 
\begin{equation}
\beta \equiv \left[ 1-\left( \frac{\Omega r}{c}\right) ^{2}\sin ^{2}\theta
\left( 1-\frac{\sin ^{2}\alpha \sin ^{2}\phi }{3\cos ^{2}\theta ^{\prime }+1}%
\right) \right] ^{1/2}-\frac{\Omega r}{c}\frac{\sin \alpha \sin \theta \sin
\phi }{\left( 3\cos ^{2}\theta ^{\prime }+1\right) ^{1/2}},  \label{beta}
\end{equation}%
with $c$ being the speed of light and $\cos \theta ^{\prime }\equiv \cos
\alpha \cos \theta +\sin \alpha \sin \theta \cos \phi $.

However, during the rotation the azimuthal phase changes by $\phi \sim
\Omega t$, while our light ray has propagated a distance by $s\sim ct$. We
describe the propagation of the light ray with the consideration of the MHD
effect above, assuming $\phi $ to be very small; e.g., $\phi \lesssim
10^{-1} $ is considered for a millisecond pulsar with $\Omega \sim 10^{2}\,%
\mathrm{Hz}$, during the time of rotation $t\lesssim 10^{-3}\,\mathrm{s}$,
such that $s\lesssim 10^{7}\,\mathrm{cm}$, which corresponds to the
propagation distance within about $10$ times the neutron star radius. Then,
for equation (\ref{mhd}) we take only the leading order expansions of $\mathbf{%
\hat{B}}\left( r_{\mathrm{o}},\theta _{\mathrm{o}},\phi \right) $ and $\beta
\left( r_{\mathrm{o}},\theta _{\mathrm{o}},\phi \right) $ in $\phi $ from
equations (\ref{B}) and (\ref{beta}), respectively, and obtain $\mathbf{\hat{n}}_{%
\left[ 0\right] }=\hat{n}_{x\left[ 0\right] }\mathbf{e}_{x}+\hat{n}_{y\left[
0\right] }\mathbf{e}_{y}+\hat{n}_{z\left[ 0\right] }\mathbf{e}_{z}$,
expressed in Cartesian coordinates: 
\begin{align}
\hat{n}_{x\left[ 0\right] }\approx \frac{2\cos \left( \theta _{\mathrm{o}%
}-\alpha \right) \sin \theta _{\mathrm{o}}+\sin \left( \theta _{\mathrm{o}%
}-\alpha \right) \cos \theta _{\mathrm{o}}}{\left( 3\cos ^{2}\left( \theta _{%
\mathrm{o}}-\alpha \right) +1\right) ^{1/2}}+\mathcal{O}\left( \phi
^{2},\left( \Omega r_{\mathrm{o}}/c\right) ^{2},\phi \left( \Omega r_{%
\mathrm{o}}/c\right) \right) ,  \label{nox} \\
\hat{n}_{z\left[ 0\right] }\approx \frac{2\cos \left( \theta _{\mathrm{o}%
}-\alpha \right) \cos \theta _{\mathrm{o}}-\sin \left( \theta _{\mathrm{o}%
}-\alpha \right) \sin \theta _{\mathrm{o}}}{\left( 3\cos ^{2}\left( \theta _{%
\mathrm{o}}-\alpha \right) +1\right) ^{1/2}}+\mathcal{O}\left( \phi
^{2},\left( \Omega r_{\mathrm{o}}/c\right) ^{2},\phi \left( \Omega r_{%
\mathrm{o}}/c\right) \right) ,  \label{noz}
\end{align}%
and 
\begin{equation}
\hat{n}_{y\left[ 0\right] }\approx \frac{\Omega }{c}\left[\frac{\sin \alpha s }{\left( 3\cos ^{2}\left( \theta _{%
\mathrm{o}}-\alpha \right) +1\right) ^{1/2}}+x_{\mathrm{o}}\right] 
+\mathcal{O}\left( \phi ^{2},\left( \Omega r_{\mathrm{o}}/c\right) ^{2},\phi
\left( \Omega r_{\mathrm{o}}/c\right) \right) ,  \label{noy}
\end{equation}%
where we have considered $\Omega r_{\mathrm{o}}/c\lesssim \phi $, e.g., for
a millisecond pulsar with $\Omega \sim 10^{2}\,\mathrm{Hz}$ and $r_{\mathrm{o%
}}\sim 10^{6}\,\mathrm{cm}$, such that $\mathcal{O}\left(\left( \Omega r_{\mathrm{o}%
}/c\right) ^{2}\right)\lesssim \mathcal{O}\left(\phi \left( \Omega r_{\mathrm{o}}/c\right)\right) \lesssim
\mathcal{O}\left(\phi ^{2}\right)$, all to be ignored in our analysis, 
and have substituted $\phi =\Omega s/c$ in equation (\ref{noy}), the leading order rotational effect to be considered in our analysis.

Now, integrating $\mathbf{\hat{n}}_{\left[ 0\right] }=\mathrm{d}\mathbf{r/}%
\mathrm{d}s$ with respect to $s$, the unperturbed (classical) trajectory of
the light ray can be derived: 
\begin{align}
& x\approx \hat{n}_{x\left[ 0\right] }s+x_{\mathrm{o}},  \label{xs} \\
& z\approx \hat{n}_{z\left[ 0\right] }s+z_{\mathrm{o}},  \label{zs} \\
& y\approx \int \hat{n}_{y\left[ 0\right] }\,\mathrm{d}s+y_{\mathrm{o}}=%
\frac{\Omega }{c}\left[ \frac{\sin \alpha s^{2}}{2\left( 3\cos ^{2}\left( \theta _{\mathrm{o}}-\alpha \right)
+1\right) ^{1/2}} +x_{\mathrm{o}}s\right]+y_{\mathrm{o}},
\label{ys}
\end{align}%
where $\hat{n}_{x\left[ 0\right] }$, $\hat{n}_{z\left[ 0\right] }$ and $\hat{%
n}_{y\left[ 0\right] }$ are given by equations (\ref{nox})-(\ref{noy}),
respectively, and the emission point is $\left(x_{\mathrm{o}},y_{\mathrm{o}%
},z_{\mathrm{o}}\right)=(r_{\mathrm{o}}\sin\theta_{\mathrm{o}},0,\allowbreak r_{%
\mathrm{o}}\cos\theta_{\mathrm{o}})$. Note that the classical
trajectory of the light ray approximates to a three-dimensional parabolic
curve in the limit $\phi \ll 1$; this results from $\hat{n}_{y\left[ 0\right]
}$ growing linearly with $s$ while $\hat{n}_{x\left[ 0\right] }$ and $\hat{n}%
_{z\left[ 0\right] }$ being constants.\footnote{%
Being projected onto the $xz$-plane, the curve appears to be a straight
line, as represented by the red dashed line in Fig. \ref{fig1}.}

In equation (\ref{n2}) $\vartheta $ must be defined as the angle between the
classical trajectory of the light ray traced by $\mathbf{\hat{n}}_{\left[ 0%
\right] }$ and the local magnetic field line $\mathbf{B}$ since $\sin
\vartheta $ is considered to be unperturbed in view of equation (\ref{n2}) (see
Fig. \ref{fig1}). Then from equations (\ref{B}) and (\ref{mhd}) one can express 
\begin{equation}
\cos \vartheta =\mathbf{\hat{n}}_{\left[ 0\right] }\cdot \mathbf{\hat{B}}%
\left( r,\theta ,\phi \right) \approx \frac{4\cos \left( \theta _{\mathrm{o}%
}-\alpha \right) \cos \left( \theta -\alpha \right) +\sin \left( \theta _{%
\mathrm{o}}-\alpha \right) \sin \left( \theta -\alpha \right) }{\left( 3\cos
^{2}\left( \theta _{\mathrm{o}}-\alpha \right) +1\right) ^{1/2}\left( 3\cos
^{2}\left( \theta -\alpha \right) +1\right) ^{1/2}}+\mathcal{O}\left( \phi
^{2},\phi \left( \Omega r_{\mathrm{o}}/c\right) \right) ,  \label{cos}
\end{equation}%
taking the leading order expansion in $\phi $. In the case of the PM
Lagrangian model, one can determine the leading order correction to $n$ by
means of equations (\ref{n2}), (\ref{B}) and (\ref{cos}): 
\begin{equation}
\delta n_{\left[ 1\right] }=\eta _{2}B^{2}\sin ^{2}\vartheta \approx \frac{%
4\eta _{2}\mu ^{2}\sin ^{2}\left( \theta -\theta _{\mathrm{o}}\right) }{%
\left( 3\cos ^{2}\left( \theta _{\mathrm{o}}-\alpha \right) +1\right) \rho
^{6}}+\mathcal{O}\left( \phi ^{2},\phi \left( \Omega r_{\mathrm{o}}/c\right)
\right) ,  \label{dn}
\end{equation}%
where $\rho \equiv \sqrt{x^{2}+z^{2}}$ with $x=\rho \sin \theta $ and $%
z=\rho \cos \theta $. For computational convenience, equation (\ref{dn}) can be
rewritten in Cartesian coordinates by substituting $\sin \theta =x/\sqrt{%
x^{2}+z^{2}}$ and $\cos \theta =z/\sqrt{x^{2}+z^{2}}$:%
\begin{equation}
\delta n_{\left[ 1\right] }\approx \frac{4\eta _{2}\mu ^{2}\left( \cos
\theta _{\mathrm{o}}x-\sin \theta _{\mathrm{o}}z\right) ^{2}}{\left( 3\cos
^{2}\left( \theta _{\mathrm{o}}-\alpha \right) +1\right) \left(
x^{2}+z^{2}\right) ^{4}},  \label{dn1}
\end{equation}

Using equation(\ref{n3}), one can easily compute the $x$ and $z$ components of $%
\delta \mathbf{\hat{n}}_{\left[ 1\right] }$:%
\begin{align}
& \delta \hat{n}_{x\left[ 1\right] }\approx \int \partial _{x}\delta n_{%
\left[ 1\right] }\,\mathrm{d}s=\frac{\delta n_{\left[ 1\right] }}{\hat{n}_{x%
\left[ 0\right] }},  \label{dnx} \\
& \delta \hat{n}_{z\left[ 1\right] }\approx \int \partial _{z}\delta n_{%
\left[ 1\right] }\,\mathrm{d}s=\frac{\delta n_{\left[ 1\right] }}{\hat{n}_{z%
\left[ 0\right] }},  \label{dnz}
\end{align}%
where $\delta n_{\left[ 1\right] }$ is given by equation (\ref{dn1}), and in
order to simplify our calculations we have exploited the relation, 
\begin{equation}
\mathrm{d}s=\mathbf{\hat{n}}_{\left[ 0\right] }\cdot \mathrm{d}\mathbf{r}=%
\hat{n}_{x\left[ 0\right] }\mathrm{d}x+\hat{n}_{z\left[ 0\right] }\mathrm{d}%
z+\hat{n}_{y\left[ 0\right] }\,\mathrm{d}y\approx \hat{n}_{x\left[ 0\right] }%
\mathrm{d}x+\hat{n}_{z\left[ 0\right] }\mathrm{d}z+\mathcal{O}\left( \left(
\Omega r_{\mathrm{o}}/c\right) ^{2}\right) \approx \frac{\mathrm{d}x}{\hat{n}%
_{x\left[ 0\right] }}~\text{or~}\frac{\mathrm{d}z}{\hat{n}_{z\left[ 0\right]
}},  \label{ds}
\end{equation}%
which is due to equations (\ref{xs})-(\ref{ys}).

To provide further convenience for equation (\ref{dn1}), we may re-parametrise
the variables $x$ and $z$ given by equations (\ref{xs}) and (\ref{zs}) using a
dimensionless parameter $\lambda \geq 0$, defined via $s=r_{\mathrm{o}%
}\lambda $:%
\begin{align}
& x=r_{\mathrm{o}}\left( \hat{n}_{x\left[ 0\right] }\lambda +\sin \theta _{%
\mathrm{o}}\right) ,  \label{xp} \\
& z=r_{\mathrm{o}}\left( \hat{n}_{z\left[ 0\right] }\lambda +\cos \theta _{%
\mathrm{o}}\right) ,  \label{zp}
\end{align}%
where $\hat{n}_{x\left[ 0\right] }$ and $\hat{n}_{z\left[ 0\right] }$ refer
to equations (\ref{nox}) and (\ref{noz}), respectively. It should be noted here
that the value of $\lambda $ is restricted by the condition $\phi \sim
\Omega t=\Omega r_{\mathrm{o}}\lambda /c\ll 1$; from this it follows that $%
\lambda \ll c/\left( \Omega r_{\mathrm{o}}\right) $. For example, for a
millisecond pulsar with $\Omega \sim 10^{2}\,\mathrm{Hz}$, we shall have $%
\lambda \ll 10^{2}$, given $r_{\mathrm{o}}\sim 10^{6}\,\mathrm{cm}$.

Inserting equations (\ref{xp}) and (\ref{zp}) into equations (\ref{dnx}) and (\ref{dnz}%
) through equation (\ref{dn1}), and substituting $s=r_{\mathrm{o}}\lambda $ in
equation (\ref{noy}), one can express $\hat{n}_{x}$, $\hat{n}_{z}$ and $\hat{n}%
_{y}$\ in terms of $\lambda $:%

\begin{align}
\hat{n}_{x}& =\hat{n}_{x\left[ 0\right] }+\delta \hat{n}_{x\left[ 1\right] }
\notag \\
& \approx \hat{n}_{x\left[ 0\right] }+\frac{4\eta _{2}B_{\mathrm{o}}^{2}\sin
^{2}\left( \theta _{\mathrm{o}}-\alpha \right) \lambda ^{2}}{\hat{n}_{x\left[
0\right] }\left( 3\cos ^{2}\left( \theta _{\mathrm{o}}-\alpha \right)
+1\right) \left[ \left( 3\cos ^{2}\left( \theta _{\mathrm{o}}-\alpha \right)
+1\right) ^{1/2}\lambda ^{2}+4\cos \left( \theta _{\mathrm{o}}-\alpha
\right) \lambda +\left( 3\cos ^{2}\left( \theta _{\mathrm{o}}-\alpha \right)
+1\right) ^{1/2}\right] ^{4}},  \label{nx}
\end{align}%

\begin{align}
\hat{n}_{z}& =\hat{n}_{z\left[ 0\right] }+\delta \hat{n}_{z\left[ 1\right] }
\notag \\
& \approx \hat{n}_{z\left[ 0\right] }+\frac{4\eta _{2}B_{\mathrm{o}}^{2}\sin
^{2}\left( \theta _{\mathrm{o}}-\alpha \right) \lambda ^{2}}{\hat{n}_{z\left[
0\right] }\left( 3\cos ^{2}\left( \theta _{\mathrm{o}}-\alpha \right)
+1\right) \left[ \left( 3\cos ^{2}\left( \theta _{\mathrm{o}}-\alpha \right)
+1\right) ^{1/2}\lambda ^{2}+4\cos \left( \theta _{\mathrm{o}}-\alpha
\right) \lambda +\left( 3\cos ^{2}\left( \theta _{\mathrm{o}}-\alpha \right)
+1\right) ^{1/2}\right] ^{4}},  \label{nz}
\end{align}%

\begin{equation}
\hat{n}_{y}=\hat{n}_{y\left[ 0\right] }\approx \frac{\Omega r_{\mathrm{o}}}{c%
}\left[\frac{\sin \alpha \lambda}{\left( 3\cos
^{2}\left( \theta _{\mathrm{o}}-\alpha \right) +1\right) ^{1/2}}
 +\sin \theta _{\mathrm{o}}\right] ,  \label{ny}
\end{equation}%
where $\hat{n}_{x\left[ 0\right] }$ and $\hat{n}_{z\left[ 0\right] }$\ refer
to equations (\ref{nox}) and (\ref{noz}), respectively, and $B_{\mathrm{o}}\equiv
\mu \left( 3\cos ^{2}\left( \theta _{\mathrm{o}}-\alpha \right) +1\right)
^{1/2}/r_{\mathrm{o}}^{3}$\ denotes the magnitude of the magnetic field at
the initial point $\left( x_{\mathrm{o}},y_{\mathrm{o}},z_{\mathrm{o}%
}\right) =\left( r_{\mathrm{o}}\sin \theta _{\mathrm{o}},0,r_{\mathrm{o}%
}\cos \theta _{\mathrm{o}}\right) $. From equations (\ref{nx})-(\ref{ny}) it is
evident that $\mathbf{\hat{n}}$ is no longer a unit vector; $\mathbf{\hat{n}%
\cdot \hat{n}}\approx 1+\mathcal{O}\left( _{\left[ 1\right] }\right) +%
\mathcal{O}\left( \left( \Omega r_{\mathrm{o}}/c\right) ^{2},_{\left[ 2%
\right] }\right) $.

Further, by integrating $\mathbf{\hat{n}}=\mathrm{d}\mathbf{r/}\mathrm{d}%
s=\left( \hat{n}_{x},\hat{n}_{y},\hat{n}_{z}\right) $ with respect to $s$,
with $\hat{n}_{x}$, $\hat{n}_{z}$ and $\hat{n}_{y}$ given by equations (\ref{nx}%
)-(\ref{ny}), one can construct a trajectory curve of the light ray:

\begin{equation}
X\equiv \int_{0}^{s}\hat{n}_{x}\,\mathrm{d}s=r_{\mathrm{o}}\int_{0}^{\lambda
}\hat{n}_{x}\,\mathrm{d}\lambda \approx r_{\mathrm{o}}\hat{n}_{x\left[ 0%
\right] }\lambda +\frac{4\eta _{2}B_{\mathrm{o}}^{2}r_{\mathrm{o}}\sin
^{2}\left( \theta _{\mathrm{o}}-\alpha \right) }{\hat{n}_{x\left[ 0\right]
}\left( 3\cos ^{2}\left( \theta _{\mathrm{o}}-\alpha \right) +1\right) }%
\mathcal{I}\left( \lambda \right) ,  \label{X}
\end{equation}

\begin{equation}
Z\equiv \int_{0}^{s}\hat{n}_{z}\,\mathrm{d}s=r_{\mathrm{o}}\int_{0}^{\lambda
}\hat{n}_{z}\,\mathrm{d}\lambda \approx r_{\mathrm{o}}\hat{n}_{z\left[ 0%
\right] }\lambda +\frac{4\eta _{2}B_{\mathrm{o}}^{2}r_{\mathrm{o}}\sin
^{2}\left( \theta _{\mathrm{o}}-\alpha \right) }{\hat{n}_{z\left[ 0\right]
}\left( 3\cos ^{2}\left( \theta _{\mathrm{o}}-\alpha \right) +1\right) }%
\mathcal{I}\left( \lambda \right) ,  \label{Z}
\end{equation}

\begin{equation}
Y\equiv \int_{0}^{s}\hat{n}_{y}\,\mathrm{d}s=r_{\mathrm{o}}\int_{0}^{\lambda}
\hat{n}_{y\left[ 0\right] }\,\mathrm{d}\lambda \approx \frac{\Omega r_{\mathrm{o}}^{2}}{c}
\left[ \frac{\sin \alpha \lambda^{2}}{2\left( 3\cos ^{2}\left( \theta _{\mathrm{o}}
-\alpha \right) +1\right) ^{1/2}} +\sin \theta _{\mathrm{o}} \lambda \right],  \label{Y}
\end{equation}%
where

\begin{align}
\mathcal{I}\left( \lambda \right) &\equiv \int_{0}^{\lambda }\frac{\lambda
^{2}}{\left[ \left( 3\cos ^{2}\left( \theta _{\mathrm{o}}-\alpha \right)
+1\right) ^{1/2}\lambda ^{2}+4\cos \left( \theta _{\mathrm{o}}-\alpha
\right) \lambda +\left( 3\cos ^{2}\left( \theta _{\mathrm{o}}-\alpha \right)
+1\right) ^{1/2}\right] ^{4}}\mathrm{d}\lambda  \notag \\
&= \frac{\left( 19\cos ^{2}\left( \theta _{\mathrm{o}}-\alpha \right)
+1\right) \left( 3\cos ^{2}\left( \theta _{\mathrm{o}}-\alpha \right)
+1\right) ^{1/2}}{16\sin ^{7}\left( \theta _{\mathrm{o}}-\alpha \right) }%
\arctan \left( \frac{\left( 3\cos ^{2}\left( \theta _{\mathrm{o}}-\alpha
\right) +1\right) ^{1/2}\lambda +2\cos \left( \theta _{\mathrm{o}}-\alpha
\right) }{\sin \left( \theta _{\mathrm{o}}-\alpha \right) }\right)  \notag \\
& \hspace{10pt} +\frac{1}{192\sin ^{6}\left( \theta _{\mathrm{o}}-\alpha \right) \left[
\left( 3\cos ^{2}\left( \theta _{\mathrm{o}}-\alpha \right) +1\right)
^{1/2}\lambda ^{2}+4\cos \left( \theta _{\mathrm{o}}-\alpha \right) \lambda
+\left( 3\cos ^{2}\left( \theta _{\mathrm{o}}-\alpha \right) +1\right) ^{1/2}%
\right] ^{3}}  \notag \\
& \hspace{20pt} \times \left[ 12\left( 3\cos ^{2}\left( \theta _{\mathrm{o}}-\alpha
\right) +1\right) ^{1/2}\left( 57\cos ^{4}\left( \theta _{\mathrm{o}}-\alpha
\right) +22\cos ^{2}\left( \theta _{\mathrm{o}}-\alpha \right) +1\right)
\lambda ^{5}\right.  \notag \\
& \left. \hspace{33pt} +120\cos \left( \theta _{\mathrm{o}}-\alpha \right) \left( 57\cos
^{4}\left( \theta _{\mathrm{o}}-\alpha \right) +22\cos ^{2}\left( \theta _{%
\mathrm{o}}-\alpha \right) +1\right) \lambda ^{4}\right.  \notag \\
& \left. \hspace{33pt} +32\left( 3\cos ^{2}\left( \theta _{\mathrm{o}}-\alpha \right)
+1\right) ^{1/2}\left( 266\cos ^{4}\left( \theta _{\mathrm{o}}-\alpha
\right) +33\cos ^{2}\left( \theta _{\mathrm{o}}-\alpha \right) +1\right)
\lambda ^{3}\right.  \notag \\
& \left. \hspace{33pt} +192\cos \left( \theta _{\mathrm{o}}-\alpha \right) \left( 76\cos
^{4}\left( \theta _{\mathrm{o}}-\alpha \right) +23\cos ^{2}\left( \theta _{%
\mathrm{o}}-\alpha \right) +1\right) \lambda ^{2}\right.  \notag \\
& \left. \hspace{33pt} +12\left( 3\cos ^{2}\left( \theta _{\mathrm{o}}-\alpha \right)
+1\right) ^{1/2}\left( 319\cos ^{4}\left( \theta _{\mathrm{o}}-\alpha
\right) +82\cos ^{2}\left( \theta _{\mathrm{o}}-\alpha \right) -1\right)
\lambda \right.  \notag \\
& \left. \hspace{33pt} +8\cos \left( \theta _{\mathrm{o}}-\alpha \right) \left( 141\cos
^{4}\left( \theta _{\mathrm{o}}-\alpha \right) +86\cos ^{2}\left( \theta _{%
\mathrm{o}}-\alpha \right) +13\right) _{\,_{{}}}^{\,^{{}}}\!\!\right]  \notag
\\
& \hspace{10pt} -\left\{ \left. \mathrm{above}\right\vert _{\lambda =0}\right\} .
\label{I}
\end{align}%
%
%
%
%
%
%
%
%
%
%
%
%
%
%
%
%
%
%
%
%
%
%

\begin{figure*}
\centering
\subfloat[]{
		\includegraphics[angle=0, height=8.7cm]{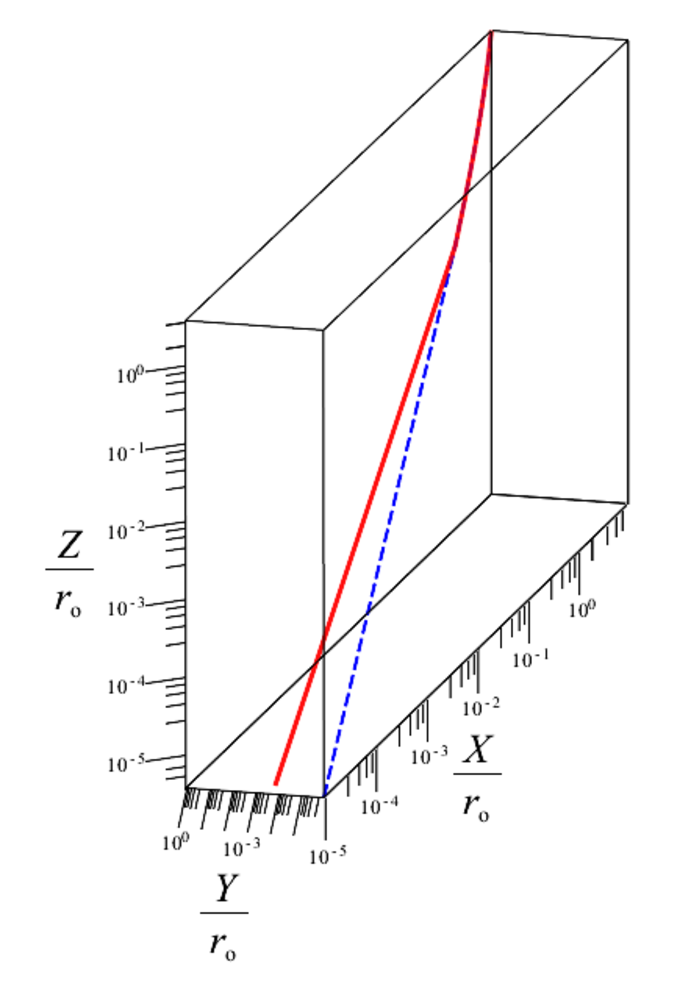}
		\label{fig:subfig01}} 
\hspace{4.2em}
\subfloat[]{
		\includegraphics[angle=0, height=9.3cm]{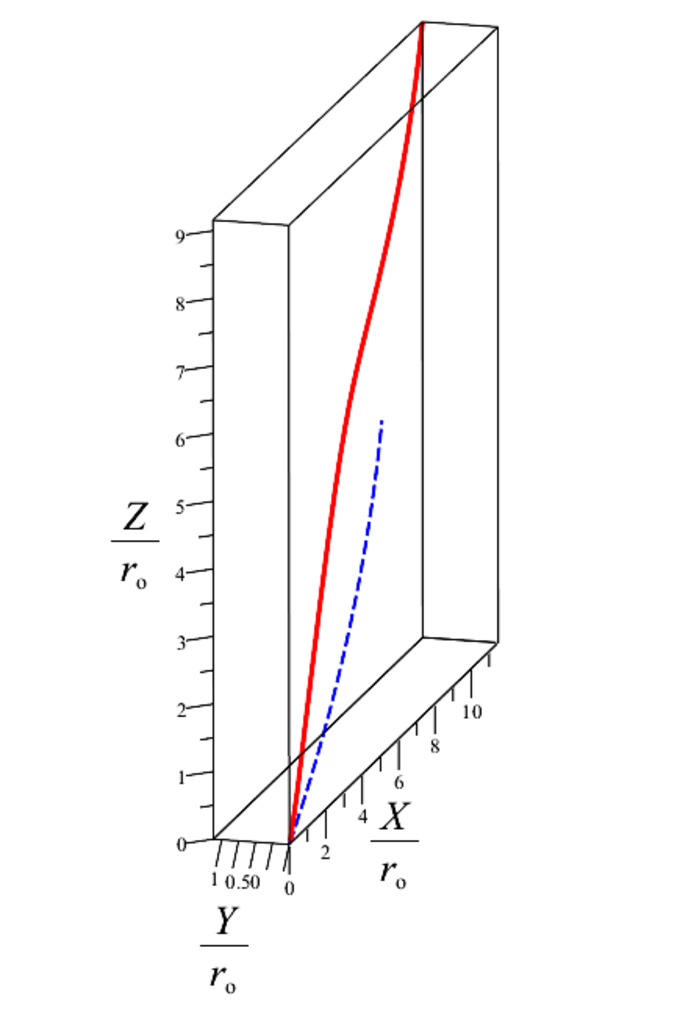}
		\label{fig:subfig02}} 
\caption{(a) A trajectory of the light ray $\left( X/r_{\mathrm{o}},Y/r_{\mathrm{%
\ o}},Z/r_{\mathrm{o}}\right) $ plotted against $0\leq \protect\lambda \leq
10$ on a logarithmic scale; the red solid curve and the blue dashed curve represent the total trajectory (classical
trajectory $+$ quantum correction) and the classical trajectory, respectively, (b) The trajectory plotted on a linear scale for intuitive visualization, 
with the quantum refraction effect fairly exaggerated by $\protect\eta _{2}B_{\mathrm{o}}^{2}\sim 10^{4}$, which is $10^{9}$ times as
large as an actual order $\sim 10^{-5}$.}
\label{fig2}
\end{figure*}

For example, with $r_{\mathrm{o}}=2\times 10^{6}\,\mathrm{cm}$, $\theta _{%
\mathrm{o}}=60^{\circ }$, $\alpha =45^{\circ }$, $\Omega =2\pi \times
10^{2}\,\mathrm{Hz}$ and $\eta _{2}B_{\mathrm{o}}^{2}\approx 4.29\times10^{-5}$,
we plot a trajectory of our light ray 
$\left( X/r_{\mathrm{o}},Y/r_{\mathrm{o}},Z/r_{\mathrm{o}}\right) $ for $%
0\leq \lambda \leq 10$ on a logarithmic scale, as shown in Fig. \ref{fig:subfig01}. 
Also, for intuitive visualization, in Fig. \ref{fig:subfig02} is plotted the trajectory on a linear scale, 
with the quantum refraction effect fairly exaggerated by $\eta _{2}B_{\mathrm{o}}^{2}\sim 10^{4}$, 
which is $10^{9}$ times as large as an actual order $\sim 10^{-5}$. Note, in particular, that the trajectory is deflected from a straight line as viewed
in the $xz$-plane (due to the quantum refraction effect), and at the same
time that it follows a parabolic path in another plane perpendicular to the $%
xz$-plane (due to the rotational effect of the pulsar magnetosphere);
therefore, the light ray follows a three-dimensional \textit{twisted} curve. In Appendix %
\ref{app} we provide a detailed discussion of the properties of this curve
in reference to the Frenet--Serret formulas (\citet{Spivak1999Comprehensive}%
).

\section{Change of polarization of a light ray due to quantum refraction}

\label{cplr}

In Section \ref{dlr} we have separated Case I and Case II for the two
different values of the refractive index $n$ attributed to the same
magnetosphere, depending on the propagation and polarization of the light
ray associated with them, as given by equation (\ref{n1}), according to the 
PM Lagrangian model (\citet{Kim202231}). In accordance
with our perturbation analysis, the refractive index can be approximated via
expansion as (\cite{Adler1971Photon})
\begin{equation}
n\approx \left\{ 
\begin{array}{ll}
1+\eta _{2}B^{2}\sin ^{2}\vartheta +\mathcal{O}%
\left( _{\left[ 2\right] }\right) & \text{for
Case I}, \\ 
1+\eta _{1}B^{2}\sin ^{2}\vartheta +\mathcal{O}%
\left( _{\left[ 2\right] }\right) & \text{for
Case II},%
\end{array}%
\right.  \label{n12}
\end{equation}%
where $\eta _{1}B^{2}$, $\eta _{2}B^{2}\lesssim 10^{-4}\ll 1$ and $\mathcal{O}%
\left( _{\left[ 2\right] }\right) $ is a shorthand expression for $\mathcal{O%
}\left( \eta _{1}^{2}B^{4},\eta _{1}\eta _{2}B^{4},\eta _{2}^{2}B^{4}\right) 
$. From this, one can see that all the results obtained through the perturbation analysis
in Section \ref{dlr} for Case I can be recycled for Case II simply by
replacing $\eta _{2}$ by $\eta _{1}$. That is, the propagation vector and
the trajectory curve of our light ray for Case II shall be given by the same
expressions as equations (\ref{nx})-(\ref{ny}) and (\ref{X})-(\ref{Y}),
respectively, but with $\eta _{2}$ replaced by $\eta _{1}$.

In this Section we work out the two polarization vectors of the light ray for Case I and Case II, associated with the dual refractive index given by (\ref{n12}); 
in contrast with the propagation of the ray, there is a distinct difference between them. In
relation to this, we discuss quantum birefringence for our pulsar emission at the end.

\subsection{For Case I\label{casei}}

According to \citet{Born1999Principles}, the propagation of the unit
polarization vector $\boldsymbol{\varepsilon}$ can be described by the
equation: 
\begin{equation}
\frac{\mathrm{d}\boldsymbol{\varepsilon}}{\mathrm{d}\tau }\equiv n\frac{\mathrm{%
d}\boldsymbol{\varepsilon}}{\mathrm{d}s}=-\left( \boldsymbol{\varepsilon}\cdot
\nabla \left( \ln n\right) \right) \nabla \mathcal{S},  \label{pol1}
\end{equation}%
where $\mathcal{S}\left( x,y,z\right) =\mathrm{const.}$ represents the
geometrical wave-front. Substituting equation (\ref{lr1}) into equation (\ref{pol1}),
we get%
\begin{equation}
\frac{\mathrm{d}\boldsymbol{\varepsilon}}{\mathrm{d}s}=-\left( \boldsymbol{\varepsilon}\cdot \nabla \left( \ln n\right) \right) \frac{\mathrm{d}%
\mathbf{r}}{\mathrm{d}s}.  \label{pol2}
\end{equation}%
Now, in view of equation (\ref{n2}), we find 
\begin{equation}
\ln n\approx \ln \left( 1+\delta n_{\left[ 1\right] }+\mathcal{O}\left( _{%
\left[ 2\right] }\right) \right) \approx \delta n_{\left[ 1\right] }+%
\mathcal{O}\left( _{\left[ 2\right] }\right) ,  \label{pol3}
\end{equation}%
where $\delta n_{\left[ 1\right] }$ refers to equation (\ref{dn}). 
Then plugging this into equation (\ref{pol2}) and inspecting the orders of
both sides, one can derive%
\begin{equation}
\frac{\mathrm{d}}{\mathrm{d}s}\left( \delta \boldsymbol{\varepsilon}_{\left[ 1%
\right] }\right) =-\left[ \boldsymbol{\varepsilon}_{\left[ 0\right] }\cdot
\nabla \left( \delta n_{\left[ 1\right] }\right) \right] \mathbf{\hat{n}}_{%
\left[ 0\right] },  \label{pol4}
\end{equation}%
where $\boldsymbol{\varepsilon}_{\left[ 0\right] }$ denotes the classical
polarization vector and $\delta \boldsymbol{\varepsilon}_{\left[ 1\right] }$ is
the leading (first) order quantum correction to it, and $\mathbf{\hat{n}}_{%
\left[ 0\right] }$ refers to the classical propagation vector. This equation
describes how quantum refraction affects the propagation of our
polarization vector along the path of the light ray by means of perturbation.

One possible way of prescribing the polarization vector classically, with the consideration of the rotational effect is 
\begin{equation}
\boldsymbol{\varepsilon}_{\left[ 0\right] }=\hat{n}_{z\left[ 0\right] }\mathbf{e%
}_{x}\pm \hat{n}_{y\left[ 0\right] }\mathbf{e}_{y}-\hat{n}_{x\left[ 0\right]
}\mathbf{e}_{z},  \label{e0}
\end{equation}%
where $\hat{n}_{x\left[ 0\right] }$, $\hat{n}_{z\left[ 0\right] }$
and $\hat{n}_{y\left[ 0\right] }$\ are given by equations (\ref{nox})-(\ref%
{noy}), respectively.\ It can be easily checked out that $\boldsymbol{\varepsilon}_{\left[ 0\right] }$ is orthogonal to the propagation vector, $%
\mathbf{\hat{n}}_{\left[ 0\right] }=\hat{n}_{x\left[ 0\right] }\mathbf{e}%
_{x}+\hat{n}_{y\left[ 0\right] }\mathbf{e}_{y}+\hat{n}_{z\left[ 0\right] }%
\mathbf{e}_{z}$, i.e., $\boldsymbol{\varepsilon}_{\left[ 0\right] }\cdot 
\mathbf{\hat{n}}_{\left[ 0\right] }\approx 0+\mathcal{O}\left( \left( \Omega
r_{\mathrm{o}}/c\right) ^{2}\right) $ while it is normalized, i.e., $\boldsymbol{\varepsilon}_{\left[ 0\right] }^{2}\approx 1+\mathcal{O}\left( \left(
\Omega r_{\mathrm{o}}/c\right) ^{2}\right) $.

Following \citet{Kim202231}, the initial polarization vector associated with 
$n$ for Case I in accordance with equation (\ref{n1}) can be expressed as%
\begin{equation}
\boldsymbol{\varepsilon}_{\mathrm{o}}=\left( 1-\eta _{1}B_{\mathrm{o}%
}^{2}+2\eta _{2}B_{\mathrm{o}}^{2}\right) \hat{n}_{z\left[ 0\right] }\mathbf{%
e}_{x}\pm \left. \hat{n}_{y\left[ 0\right] }\right\vert _{s=0}\mathbf{e}%
_{y}-\left( 1-\eta _{1}B_{\mathrm{o}}^{2}\right) \hat{n}_{x\left[ 0\right] }%
\mathbf{e}_{z},  \label{eo}
\end{equation}%
which has been adapted from its original expression in \citet{Kim202231} to
the geometry of our rotating magnetosphere, with the consideration of equation (%
\ref{e0}).\footnote{%
Originally, in \citet{Kim202231} the initial polarization vector associated
with $n$ for Case I is given by $\boldsymbol{\varepsilon}_{\mathrm{o}}=\left(
1-\eta _{1}B_{\mathrm{o}}^{2}+2\eta _{2}B_{\mathrm{o}}^{2}\right) \cos
\theta \mathbf{e}_{x}-\left( 1-\eta _{1}B_{\mathrm{o}}^{2}\right) \sin
\theta \mathbf{e}_{z}$, which corresponds to the classical polarization
vector, $\boldsymbol{\varepsilon}_{\left[ 0\right] }=$ $\cos \theta \mathbf{e}%
_{x}-\sin \theta \mathbf{e}_{z}$.} Then we may separate the classical part, 
\begin{equation}
\boldsymbol{\varepsilon}_{\mathrm{o}\left[ 0\right] }\equiv \left. \boldsymbol{\varepsilon}_{\left[ 0\right] }\right\vert _{s=0}=\hat{n}_{z\left[ 0\right]
}\mathbf{e}_{x}\pm \left. \hat{n}_{y\left[ 0\right] }\right\vert _{s=0}%
\mathbf{e}_{y}-\hat{n}_{x\left[ 0\right] }\mathbf{e}_{z}  \label{eo0}
\end{equation}%
and the quantum correction, 
\begin{equation}
\delta \boldsymbol{\varepsilon}_{\mathrm{o}\left[ 1\right] }=\left( -\eta
_{1}B_{\mathrm{o}}^{2}+2\eta _{2}B_{\mathrm{o}}^{2}\right) \hat{n}_{z\left[ 0%
\right] }\mathbf{e}_{x}+\eta _{1}B_{\mathrm{o}}^{2}\hat{n}_{x\left[ 0\right]
}\mathbf{e}_{z}.  \label{eo1}
\end{equation}

The polarization vector with the first order correction due to the quantum
refraction effect can be obtained in a similar manner as in Sect \ref{dlr}.
Integrating equation (\ref{pol4}) with respect to $s=r_{\mathrm{o}}\lambda $,
and\ combining this with equation (\ref{e0}), and using equations (\ref{dn1}), (\ref%
{ds}), (\ref{xp}), (\ref{zp}), (\ref{e0}) and (\ref{eo1}), we finally have $%
\boldsymbol{\varepsilon}=\varepsilon _{x}\mathbf{e}_{x}+\varepsilon _{y}\mathbf{%
e}_{y}+\varepsilon _{z}\mathbf{e}_{z}$ with 

\begin{align}
\varepsilon _{x}&= \varepsilon _{x\left[ 0\right] }+\delta \varepsilon _{x%
\left[ 1\right] }  \notag \\
&\approx \hat{n}_{z\left[ 0\right] }+\left( -\eta _{1}B_{\mathrm{o}%
}^{2}+2\eta _{2}B_{\mathrm{o}}^{2}\right) \hat{n}_{z\left[ 0\right] }  \notag
\\
&\hspace{10pt} -\frac{4\eta _{2}B_{\mathrm{o}}^{2}\sin ^{2}\left( \theta _{\mathrm{o}%
}-\alpha \right) \left( \hat{n}_{z\left[ 0\right] }^{2}-\hat{n}_{x\left[ 0%
\right] }^{2}\right) \lambda ^{2}}{\hat{n}_{z\left[ 0\right] }\left( 3\cos
^{2}\left( \theta _{\mathrm{o}}-\alpha \right) +1\right) \left[ \left( 3\cos
^{2}\left( \theta _{\mathrm{o}}-\alpha \right) +1\right) ^{1/2}\lambda
^{2}+4\cos \left( \theta _{\mathrm{o}}-\alpha \right) \lambda +\left( 3\cos
^{2}\left( \theta _{\mathrm{o}}-\alpha \right) +1\right) ^{1/2}\right] ^{4}},
\label{ex}
\end{align}%

\begin{align}
\varepsilon _{z}&= \varepsilon _{z\left[ 0\right] }+\delta \varepsilon _{z%
\left[ 1\right] }  \notag \\
&\approx -\hat{n}_{x\left[ 0\right] }+\eta _{1}B_{\mathrm{o}}^{2}\hat{n}_{x%
\left[ 0\right] }  \notag \\
&\hspace{10pt} -\frac{4\eta _{2}B_{\mathrm{o}}^{2}\sin ^{2}\left( \theta _{\mathrm{o}%
}-\alpha \right) \left( \hat{n}_{z\left[ 0\right] }^{2}-\hat{n}_{x\left[ 0%
\right] }^{2}\right) \lambda ^{2}}{\hat{n}_{x\left[ 0\right] }\left( 3\cos
^{2}\left( \theta _{\mathrm{o}}-\alpha \right) +1\right) \left[ \left( 3\cos
^{2}\left( \theta _{\mathrm{o}}-\alpha \right) +1\right) ^{1/2}\lambda
^{2}+4\cos \left( \theta _{\mathrm{o}}-\alpha \right) \lambda +\left( 3\cos
^{2}\left( \theta _{\mathrm{o}}-\alpha \right) +1\right) ^{1/2}\right] ^{4}},
\label{ez}
\end{align}%

\begin{align}
\varepsilon _{y}&= \varepsilon _{y\left[ 0\right] }+\delta \varepsilon _{y%
\left[ 1\right] }  \notag \\
&\approx \pm \hat{n}_{y\left[ 0\right] }-\frac{4\eta _{2}B_{\mathrm{o}%
}^{2}\sin ^{2}\left( \theta _{\mathrm{o}}-\alpha \right) \left( \hat{n}_{z%
\left[ 0\right] }^{2}-\hat{n}_{x\left[ 0\right] }^{2}\right) }{\hat{n}_{x%
\left[ 0\right] }\hat{n}_{z\left[ 0\right] }\left( 3\cos ^{2}\left( \theta _{%
\mathrm{o}}-\alpha \right) +1\right) }\left[ \frac{\hat{n}_{y\left[ 0\right]
}\lambda ^{2}}{\left[ \left( 3\cos ^{2}\left( \theta _{\mathrm{o}}-\alpha
\right) +1\right) ^{1/2}\lambda ^{2}+4\cos \left( \theta _{\mathrm{o}%
}-\alpha \right) \lambda +\left( 3\cos ^{2}\left( \theta _{\mathrm{o}%
}-\alpha \right) +1\right) ^{1/2}\right] ^{4}}\right.   \notag \\
\!& \left. \left. \hspace{8pt} - \frac{\Omega r_{\mathrm{o}}\sin \alpha }{c\left( 3\cos ^{2}\left( \theta _{\mathrm{o}%
}-\alpha \right) +1\right) ^{1/2}} \mathcal{I}\left( \lambda \right)
\right. _{\,_{\,_{\,_{\,_{{}}}}}}^{\,^{\,^{\,^{\,^{{}}}}}}\!\!\!\!\!\right] ,
\label{ey}
\end{align}%
where $\hat{n}_{x\left[ 0\right] }$, $\hat{n}_{z\left[ 0\right] }$ and $\hat{%
n}_{y\left[ 0\right] }$\ are given by equations (\ref{nox}), (\ref{noz}) and (\ref%
{ny}), respectively, and $B_{\mathrm{o}}=\mu \left( 3\cos ^{2}\left( \theta
_{\mathrm{o}}-\alpha \right) +1\right) ^{1/2}/r_{\mathrm{o}}^{3}$, and $%
\mathcal{I}\left( \lambda \right) $ refers to equation (\ref{I}). In Fig. \ref%
{fig3} is plotted the change in the polarization vector $\left( \Delta
\varepsilon _{x},\Delta \varepsilon _{y},\Delta \varepsilon _{z}\right)
\equiv \left. \left( \varepsilon _{x}\left( \lambda \right) ,\varepsilon
_{y}\left( \lambda \right) ,\varepsilon _{z}\left( \lambda \right) \right)
\right\vert _{0}^{\lambda }$ against $0\leq \lambda \leq 10$ on a logarithmic scale, wherein $r_{%
\mathrm{o}}=2\times 10^{6}\,\mathrm{cm}$, $\theta _{\mathrm{o}}=60^{\circ }$%
, $\alpha =45^{\circ }$, $\Omega =2\pi \times 10^{2}\,\mathrm{Hz}$, $\eta
_{1}B_{\mathrm{o}}^{2}\approx 2.45\times10^{-5}$ and $\eta _{2}B_{\mathrm{o}}^{2}\approx 4.29\times10^{-5}$. 
From this plot, one can see that the total polarization vector changes drastically along the $x$-axis and $z$-axis 
near the beginning of the propagation of our light ray due to the quantum refraction effect,
while the classical polarization vector changes only along the $y$-axis due to the rotational effect, as can be seen from equation (\ref{e0}).

\begin{figure}
\centering
\includegraphics[height=9.5cm]{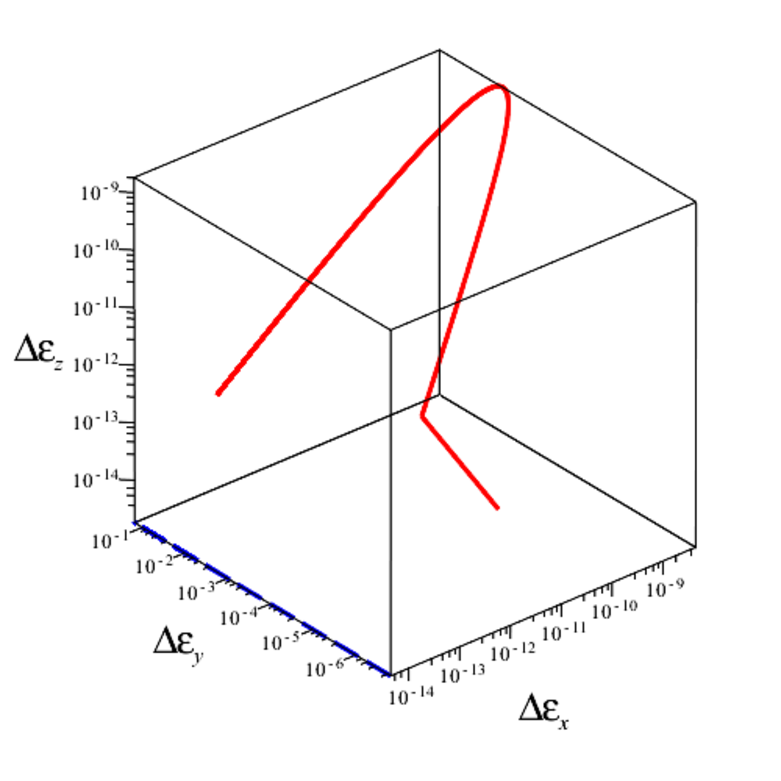}
\caption{The change in the polarization vector, $\left( \Delta \protect%
\varepsilon _{x},\Delta \protect\varepsilon _{y},\Delta \protect\varepsilon %
_{z}\right) \equiv \left. \left( \protect\varepsilon _{x}\left( \protect%
\lambda \right) ,\protect\varepsilon _{y}\left( \protect\lambda \right) ,%
\protect\varepsilon _{z}\left( \protect\lambda \right) \right) \right\vert
_{0}^{\protect\lambda }$ plotted against $0\leq \protect\lambda \leq 10$ on a logarithmic scale;
the red solid curve and the blue dashed curve represent the total polarization vector (classical
polarization vector $+$ quantum correction) and the classical polarization vector, respectively.}
\label{fig3}
\end{figure}

By equations (\ref{nx})-(\ref{ny}) and (\ref{ex})-(\ref{ey}) one can inspect the
orthogonality between the propagation and polarization vectors,$\mathbf{\ 
\hat{n}}$ and $\boldsymbol{\varepsilon}$:%
\begin{equation}
\mathbf{\hat{n}}\cdot \boldsymbol{\varepsilon}\approx 2\eta _{2}B_{\mathrm{o}%
}^{2}\hat{n}_{x\left[ 0\right] }\hat{n}_{z\left[ 0\right] }+\mathcal{O}%
\left( \left( \Omega r_{\mathrm{o}}/c\right) ^{2},_{\left[ 2\right] }\right)
,  \label{og}
\end{equation}%
where $\hat{n}_{x\left[ 0\right] }$ and $\hat{n}_{z\left[ 0\right] }$ are
given by equations (\ref{nox}) and (\ref{noz}), respectively. This implies that
the quantum refraction effect results in breaking the orthogonality at the leading
order in $\eta _{2}B_{\mathrm{o}}^{2}$. However, the departure from the
orthogonality remains constant under this effect, being determined at the
leading order in $\eta _{2}B_{\mathrm{o}}^{2}$ solely by the initial
conditions for quantum refraction. For example, with $r_{\mathrm{o}%
}=2\times 10^{6}\,\mathrm{cm}$, $\theta _{\mathrm{o}}=60^{\circ }$, $\alpha
=45^{\circ }$ and a usual value of $\eta _{2}B_{\mathrm{o}}^{2}\approx
4.29\times 10^{-5}$, we find the departure value to be $\sin^{-1}\left(\mathbf{\hat{n}}\cdot \boldsymbol{\varepsilon}\right)\approx
3.02\times 10^{-5}\,\mathrm{rad}$. 

Using equations (\ref{e0}) and (\ref{ex})-(\ref{ey}), the Faraday rotation angle
can be determined via 
\begin{equation}
\varphi _{\mathrm{F}}\equiv \cos ^{-1}\left( \boldsymbol{\varepsilon}_{\left[ 0%
\right] }\cdot \boldsymbol{\hat{\varepsilon}}\right) \approx \sqrt{\delta 
\boldsymbol{\varepsilon}_{\left[ 1\right] }^{2}-\left( \boldsymbol{\varepsilon}_{%
\left[ 0\right] }\cdot \delta \boldsymbol{\varepsilon}_{\left[ 1\right]
}\right) ^{2}}+\mathcal{O}\left( \left( \Omega r_{\mathrm{o}}/c\right)
^{2},_{\left[ 2\right] }\right) .  \label{fr}
\end{equation}%
For example, with $r_{\mathrm{o}}=2\times 10^{6}\,\mathrm{cm}$, $\theta _{%
\mathrm{o}}=60^{\circ }$, $\alpha =45^{\circ }$ and usual values of $\eta
_{1}B_{\mathrm{o}}^{2}\approx 2.45\times10^{-5}$, $\eta _{2}B_{\mathrm{o}}^{2}\approx 4.29\times10^{-5}$, 
we find $\varphi _{\mathrm{F}}\approx 3.02\times 10^{-5}\,\mathrm{rad}$.

Note that the values, $\sin^{-1}\left(\mathbf{\hat{n}}\cdot \boldsymbol{\varepsilon}\right)\approx
3.02\times 10^{-5}\,\mathrm{rad}$ and $\varphi _{\mathrm{F}}\approx 3.02\times
10^{-5}\,\mathrm{rad}$, obtained as above are small but comparable to our
perturbation parameters, $\eta
_{1}B_{\mathrm{o}}^{2}\approx 2.45\times10^{-5}$ and $\eta _{2}B_{\mathrm{o}}^{2}\approx 4.29\times10^{-5}$%
; therefore, these effects are truly quantum refractive.

\subsection{For Case II\label{caseii}}

One can write down the classical polarization vector associated with the refractive index $n$ for Case II as given by (\ref{n1}), 
with the consideration of the rotational effect as 
\begin{equation}
\boldsymbol{\varepsilon}_{\left[ 0\right] }=-\left(\hat{n}_{x\left[ 0\right] }+\hat{n}_{z\left[ 0\right] }\right) \hat{n}_{y\left[ 0\right] }
\mathbf{e}_{x}+\mathbf{e}_{y}+\left(\hat{n}_{x\left[ 0\right] }-\hat{n}_{z\left[ 0\right] }\right) \hat{n}_{y\left[ 0\right] }\mathbf{e}_{z},
\label{e00}
\end{equation}
where $\hat{n}_{x\left[ 0\right] }$, $\hat{n}_{z\left[ 0\right] }$ and $\hat{n}_{y\left[ 0\right] }$\ are given by equations (\ref{nox})-(\ref{noy}), respectively.
In accordance with this, the initial polarization vector shall be 
\begin{equation}
\boldsymbol{\varepsilon}_{\mathrm{o}}=-\left(\hat{n}_{x\left[ 0\right] }+\hat{n}_{z\left[ 0\right] }\right)\left. \hat{n}_{y\left[ 0\right] }\right\vert _{s=0}
\mathbf{e}_{x}+\mathbf{e}_{y}+\left(\hat{n}_{x\left[ 0\right] }-\hat{n}_{z\left[ 0\right] }\right)\left. \hat{n}_{y\left[ 0\right] }\right\vert _{s=0}\mathbf{e}_{z},  \label{eoo}
\end{equation}
which has been adapted from its original expression in \citet{Kim202231} to the geometry of our rotating magnetosphere.\footnote{In \citet{Kim202231} 
the initial polarization vector associated with $n$ for Case II is given by $\boldsymbol{\varepsilon}_{\mathrm{o}}=\mathbf{e}_{y}$, 
which is identical to the classical polarization vector, $\boldsymbol{\varepsilon }_{\left[ 0\right] }=\mathbf{e}_{y}$.} 

It can be checked out that $\boldsymbol{\varepsilon }_{\left[ 0\right] }$ is
orthogonal to the propagation vector, $\mathbf{\hat{n}}_{\left[ 0\right] }=%
\hat{n}_{x\left[ 0\right] }\mathbf{e}_{x}+\hat{n}_{y\left[ 0\right] }\mathbf{%
e}_{y}+\hat{n}_{z\left[ 0\right] }\mathbf{e}_{z}$, i.e., $\boldsymbol{%
\varepsilon }_{\left[ 0\right] }\cdot \mathbf{\hat{n}}_{\left[ 0\right]
}= 0$ while it is normalized, i.e., $\boldsymbol{\varepsilon }_{\left[
0\right] }^{2}\approx 1+\mathcal{O}\left( \left( \Omega r_{\mathrm{o}%
}/c\right) ^{2}\right) $. In addition, from equations (\ref{e0}) and (\ref{e00}) we find that $\boldsymbol{\varepsilon }_{\left[ 0\right]\,\left(\mathrm{Case~I}\right) }\cdot
\boldsymbol{\varepsilon }_{\left[ 0\right]\,\left(\mathrm{Case~II}\right) }=0$. Then one can note that the three vectors, $\mathbf{\hat{n}}_{\left[ 0\right] }$, $\boldsymbol{\ \varepsilon }_{\left[ 0\right]\,\left(\mathrm{Case~I}\right) }$ and 
$\boldsymbol{\varepsilon }_{\left[ 0\right]\,\left(\mathrm{Case~II}\right) }$ form a classical orthonormal basis.\footnote
{From Appendix \ref{app} one can see that $\mathbf{\hat{n}}_{\left[ 0\right] }=\mathbf{T}_{\left[ 0\right] }=\hat{n}_{x\left[ 0\right] }\mathbf{e}_{x}+\hat{n}_{y\left[ 0\right] }\mathbf{e}_{y}+\hat{n}_{z\left[ 0\right] }\mathbf{e}_{z}$, 
$\boldsymbol{\varepsilon }_{\left[ 0\right]\,\left(\mathrm{Case~I}\right) }=\mathbf{N}_{\left[ 0\right] }=\hat{n}_{z\left[ 0\right] }
\mathbf{e}_{x}+\hat{n}_{y\left[ 0\right] }\mathbf{e}_{y}-\hat{n}_{x\left[ 0\right]}\mathbf{e}_{z}$ and 
$\boldsymbol{\varepsilon }_{\left[ 0\right]\,\left(\mathrm{Case~II}\right) }=\mathbf{B}_{\left[ 0\right] }=-\left(\hat{n}_{x\left[ 0\right] }+\hat{n}_{z\left[ 0\right] }\right)\hat{n}_{y\left[ 0\right] }\mathbf{e}_{x}+\mathbf{e}_{y}+\left(\hat{n}_{x\left[ 0\right] }-\hat{n}_{z\left[ 0\right] }\right)
\hat{n}_{y\left[ 0\right] }\mathbf{e}_{z}$, where $\mathbf{T}_{\left[ 0\right] }$, $\mathbf{N}_{\left[ 0\right] }$ 
and $\mathbf{B}_{\left[ 0\right] }$ denote the unperturbed (classical)
part of the unit tangent, normal and bi-normal vectors, taken from equations (\ref{A2})-(\ref{A4}), respectively, with all the terms of $%
\delta n_{\left[ 1\right] }$ removed.}

The polarization vector with the first order correction due to the quantum
refraction effect can be obtained in the same manner as in Case I above.
That is, we combine the classical polarization vector $\boldsymbol{\varepsilon}%
_{\left[ 0\right] }$ (given by equation (\ref{e00})) and the quantum correction
(given by the integral of equation (\ref{pol4}) with respect to $s=r_{\mathrm{o}%
}\lambda $, wherein the refractive index is expressed with $\eta _{1}$,
following equation (\ref{n12})). We obtain $\boldsymbol{\varepsilon}=\varepsilon _{x}%
\mathbf{e}_{x}+\varepsilon _{y}\mathbf{e}_{y}+\varepsilon _{z}\mathbf{e}_{z}$
with 

\begin{align}
\varepsilon _{x}& = \varepsilon _{x\left[ 0\right] }+\delta \varepsilon _{x%
\left[ 1\right] }  \notag \\
& \approx -\left(\hat{n}_{x\left[ 0\right] }+\hat{n}_{z\left[ 0\right] }\right)\hat{n}_{y\left[ 0\right] }
+\frac{8\eta _{1}B_{\mathrm{o}}^{2}\sin ^{2}\left( \theta _{\mathrm{o}}-\alpha
\right) }{3\cos ^{2}\left( \theta _{\mathrm{o}}-\alpha \right) +1}\left[ 
\frac{\left(\hat{n}_{x\left[ 0\right] }+\hat{n}_{z\left[ 0\right] }\right)\hat{n}_{y\left[ 0\right] }\lambda ^{2}}{%
\left[ \left( 3\cos ^{2}\left( \theta _{\mathrm{o}}-\alpha \right) +1\right)
^{1/2}\lambda ^{2}+4\cos \left( \theta _{\mathrm{o}}-\alpha \right) \lambda
+\left( 3\cos ^{2}\left( \theta _{\mathrm{o}}-\alpha \right) +1\right) ^{1/2}%
\right] ^{4}}\right.  \notag \\
& \!\left. \left. \hspace{10pt} -\frac{\Omega r_{\mathrm{o}}\left(\hat{n}_{x\left[ 0\right] }+\hat{n}_{z\left[ 0\right] }\right)\sin \alpha}
{c\left( 3\cos^{2}\left( \theta _{\mathrm{o}}-\alpha \right) +1\right) ^{1/2}} \mathcal{I}\left( \lambda \right) \right.
_{\,_{\,_{\,_{\,_{{}}}}}}^{\,^{\,^{\,^{\,^{{}}}}}}\!\!\!\!\!\right] ,
\label{ex2}
\end{align}%

\begin{align}
\varepsilon _{z}&= \varepsilon _{z\left[ 0\right] }+\delta \varepsilon _{z%
\left[ 1\right] }  \notag \\
&\approx \left(\hat{n}_{x\left[ 0\right] }-\hat{n}_{z\left[ 0\right] }\right)\hat{n}_{y\left[ 0\right] }+\frac{%
8\eta _{1}B_{\mathrm{o}}^{2}\sin ^{2}\left( \theta _{\mathrm{o}}-\alpha
\right) }{3\cos ^{2}\left( \theta _{\mathrm{o}}-\alpha \right) +1}\left[ 
\frac{-\left(\hat{n}_{x\left[ 0\right] }-\hat{n}_{z\left[ 0\right] }\right)\hat{n}_{y\left[ 0\right] }\lambda ^{2}}{%
\left[ \left( 3\cos ^{2}\left( \theta _{\mathrm{o}}-\alpha \right) +1\right)
^{1/2}\lambda ^{2}+4\cos \left( \theta _{\mathrm{o}}-\alpha \right) \lambda
+\left( 3\cos ^{2}\left( \theta _{\mathrm{o}}-\alpha \right) +1\right) ^{1/2}%
\right] ^{4}}\right.  \notag \\
& \!\left. \left. \hspace{10pt} +\frac{\Omega r_{\mathrm{o}}\left(\hat{n}_{x\left[ 0\right] }-\hat{n}_{z\left[ 0\right] }\right)\sin \alpha}
{c\left( 3\cos^{2}\left( \theta _{\mathrm{o}}-\alpha \right) +1\right) ^{1/2}} \mathcal{I}\left( \lambda \right) \right.
_{\,_{\,_{\,_{\,_{{}}}}}}^{\,^{\,^{\,^{\,^{{}}}}}}\!\!\!\!\!\right] ,
\label{ez2}
\end{align}%

\begin{equation}
\varepsilon _{y}=\varepsilon _{y\left[ 0\right] }+\delta \varepsilon _{y%
\left[ 1\right] }\approx 1+\mathcal{O}\left( \left( \Omega r_{\mathrm{o}%
}/c\right) ^{2}\right) ,  \label{ey2}
\end{equation}%
where $\hat{n}_{x\left[ 0\right] }$, $\hat{n}_{z\left[ 0\right] }$ and $\hat{%
n}_{y\left[ 0\right] }$\ are given by equations (\ref{nox}), (\ref{noz}) and (\ref%
{ny}), respectively, and $B_{\mathrm{o}}=\mu \left( 3\cos ^{2}\left( \theta _{%
\mathrm{o}}-\alpha \right) +1\right) ^{1/2}/r_{\mathrm{o}}^{3}$, and $%
\mathcal{I}\left( \lambda \right) $ refers to equation (\ref{I}). In Fig. \ref%
{fig3.2} is plotted the change in the polarization vector $\left( \Delta
\varepsilon _{x},\Delta \varepsilon _{y},\Delta \varepsilon _{z}\right)
\equiv \left. \left( \varepsilon _{x}\left( \lambda \right) ,\varepsilon
_{y}\left( \lambda \right) ,\varepsilon _{z}\left( \lambda \right) \right)
\right\vert _{0}^{\lambda }$ against $0\leq \lambda \leq 10$ on a logarithmic scale, wherein $r_{%
\mathrm{o}}=2\times 10^{6}\,\mathrm{cm}$, $\theta _{\mathrm{o}}=60^{\circ }$%
, $\alpha =45^{\circ }$, $\Omega =2\pi \times 10^{2}\,\mathrm{Hz}$ and $\eta
_{1}B_{\mathrm{o}}^{2}\approx 2.45\times10^{-5}$. From this plot, one can see virtually no difference 
between the total polarization vector and the classical polarization vector; practically, no quantum refraction effect exists. 
However, the polarization vector changes along the $x$-axis and $z$-axis due to the rotational effect, as can be seen from equation (\ref{e00}).

\begin{figure}
\centering\includegraphics[height=9.5cm]{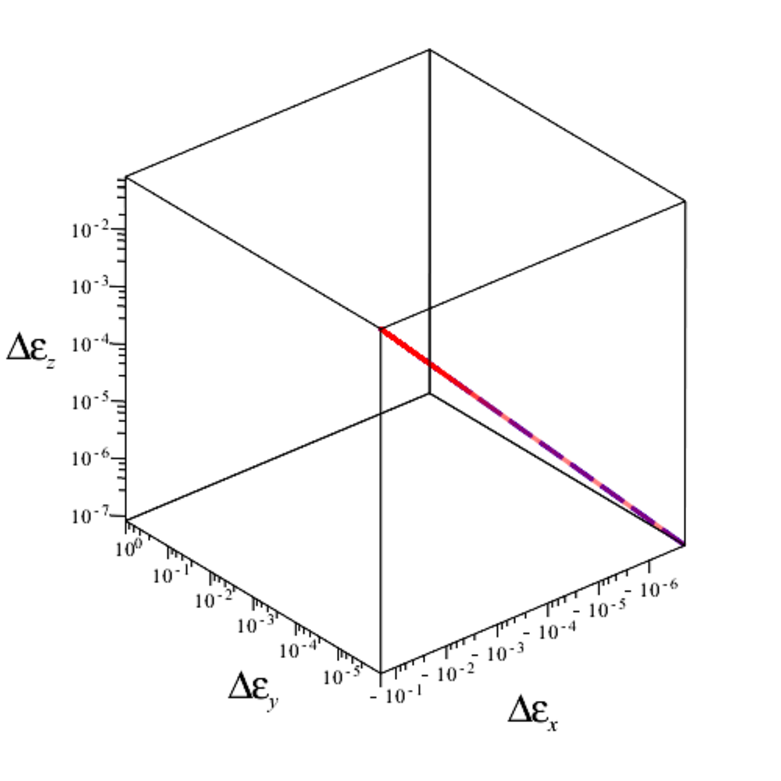}
\caption{The change in the polarization vector, $\left( \Delta \protect%
\varepsilon _{x},\Delta \protect\varepsilon _{y},\Delta \protect\varepsilon %
_{z}\right) \equiv \left. \left( \protect\varepsilon _{x}\left( \protect%
\lambda \right) ,\protect\varepsilon _{y}\left( \protect\lambda \right) ,%
\protect\varepsilon _{z}\left( \protect\lambda \right) \right) \right\vert
_{0}^{\protect\lambda }$ plotted against $0\leq \protect\lambda \leq 10$ on a logarithmic scale;
the red solid curve and the blue dashed curve represent the total polarization vector (classical
polarization vector $+$ quantum correction) and the classical polarization vector, respectively.}
\label{fig3.2}
\end{figure}

The orthogonality between the propagation and polarization vectors,$\mathbf{%
\ \hat{n}}$ and $\boldsymbol{\varepsilon}$ can be inspected using equations (\ref{nx}%
)-(\ref{ny}) and (\ref{ex2})-(\ref{ey2}):%
\begin{align}
\mathbf{\hat{n}}\cdot \boldsymbol{\varepsilon}
&\approx \frac{4\eta _{1}B_{\mathrm{o}}^{2}\sin
^{2}\left( \theta _{\mathrm{o}}-\alpha \right)\left(\hat{n}_{x\left[
0\right]}^{2}-\hat{n}_{z\left[0\right]}^{2} \right)\hat{n}_{y\left[0\right]} \lambda ^{2}}
{\hat{n}_{x\left[0\right] }\hat{n}_{z\left[0\right] }\left( 3\cos ^{2}\left( \theta _{\mathrm{o}}-\alpha \right)
+1\right) \left[ \left( 3\cos ^{2}\left( \theta _{\mathrm{o}}-\alpha \right)
+1\right) ^{1/2}\lambda ^{2}+4\cos \left( \theta _{\mathrm{o}}-\alpha
\right) \lambda +\left( 3\cos ^{2}\left( \theta _{\mathrm{o}}-\alpha \right)
+1\right) ^{1/2}\right] ^{4}}  \notag \\
&\hspace{10pt}-\frac{8\eta _{1}B_{\mathrm{o}}^{2}\Omega r_{\mathrm{o}}\sin \alpha \sin ^{2}\left( \theta _{\mathrm{o}}
-\alpha \right) }{c\left( 3\cos ^{2}\left( \theta _{\mathrm{o}}-\alpha
\right) +1\right)^{3/2} } \mathcal{I}\left( \lambda \right) +\mathcal{O}\left( \left( \Omega
r_{\mathrm{o}}/c\right) ^{2},_{\left[ 2\right] }\right) ,  \label{og2}
\end{align}%
where $\hat{n}_{x\left[ 0\right] }$, $\hat{n}_{z\left[ 0\right] }$ and $\hat{%
n}_{y\left[ 0\right] }$\ are given by equations (\ref{nox}), (\ref{noz}) and (\ref%
{ny}), respectively, and $%
\mathcal{I}\left( \lambda \right) $ refers to equation (\ref{I}). This implies
that the quantum refraction effect results in breaking the orthogonality at the
leading order in $\eta _{1}B_{\mathrm{o}}^{2}$. Note here that unlike Case I
the departure from the orthogonality changes over the propagation of the
light ray under this effect. For example, with $r_{\mathrm{o}}=2\times 10^{6}\,\mathrm{cm}$, $%
\theta _{\mathrm{o}}=60^{\circ }$, $\alpha =45^{\circ }$, $\Omega =2\pi
\times 10^{2}\,\mathrm{Hz}$ and a usual value of $\eta _{1}B_{\mathrm{o}%
}^{2}\approx 2.45\times 10^{-5}$, we find the maximum departure value to be
$\sin^{-1}\left(\mathbf{\hat{n}}\cdot \boldsymbol{\varepsilon}\right)\approx
1.05\times 10^{-10}\,\mathrm{rad}$. 

Using equations (\ref{e00}) and (\ref{ex2})-(\ref{ey2}), the Faraday rotation
angle can be determined via 
\begin{equation}
\varphi _{\mathrm{F}}=\cos ^{-1}\left( \boldsymbol{\varepsilon}_{\left[ 0\right]
}\cdot \boldsymbol{\hat{\varepsilon}}\right) \approx \sqrt{\delta \boldsymbol{%
\varepsilon }_{\left[ 1\right] }^{2}-\left( \boldsymbol{\varepsilon}_{\left[ 0%
\right] }\cdot \delta \boldsymbol{\varepsilon}_{\left[ 1\right] }\right) ^{2}}+%
\mathcal{O}\left( \left( \Omega r_{\mathrm{o}}/c\right) ^{2},_{\left[ 2%
\right] }\right) .  \label{fr2}
\end{equation}%
For example, with $r_{\mathrm{o}}=2\times 10^{6}\,\mathrm{cm}$, $\theta _{%
\mathrm{o}}=60^{\circ }$, $\alpha =45^{\circ }$, $\Omega =2\pi \times
10^{2}\,\mathrm{Hz}$ and a usual value of $\eta _{1}B_{\mathrm{o}%
}^{2}\approx 2.45\times 10^{-5}$, we find $\varphi _{\mathrm{F}}\approx
1.74\times 10^{-10}\,\mathrm{rad}$.

Note that the values, $\sin^{-1}\left(\mathbf{\hat{n}}\cdot \boldsymbol{\varepsilon}\right)\approx
1.05\times 10^{-10}\,\mathrm{rad}$ and $\varphi _{\mathrm{F}}\approx
1.74\times 10^{-10}\,\mathrm{rad}$, obtained as above are extremely small
compared to our perturbation parameter $\eta _{1}B_{\mathrm{o}}^{2}\approx
2.45\times 10^{-5}$; therefore, these quantum refraction effects can be
considered practically negligible.

\subsection{Quantum birefringence}

From equations (\ref{og}) and (\ref{og2}) above, one can note the following: the
polarization vector partly has a longitudinal component (i.e., a component
parallel to the propagation vector) for Case I, whereas it is substantially
perpendicular to the propagation vector for Case II. This is because given
the conditions for pulsar emission as above, we have $\sin^{-1}\left(\mathbf{\hat{n}}\cdot \boldsymbol{%
\varepsilon }_{\left(\mathrm{Case~I}\right)}\right)\approx 3.02\times 10^{-5}\,\mathrm{rad}$,
which is small but comparable to the perturbation parameter $\eta _{2}B_{\mathrm{o}}^{2}\approx 4.29\times 10^{-5}$, 
and therefore not negligible,
while $\sin^{-1}\left(\mathbf{\hat{n}}\cdot \boldsymbol{\varepsilon}_{\left(\mathrm{Case~II}\right)}\right)
\approx 1.05\times 10^{-10}\,\mathrm{rad}$ is practically negligible
compared to the perturbation parameter $\eta _{1}B_{\mathrm{o}}^{2}\approx
2.45\times 10^{-5}$. The two different polarization
modes, together with the dual refractive index $n$ as given by (\ref{n12}),
are entirely due to the quantum refraction effect. These optical properties
can be considered to define `quantum birefringence' as the phenomenology
involved is analogous to classical birefringence.

Classically, birefringence is a well-known phenomenon in crystal optics, but the quantum birefringence considered here has a notable difference from the crystal birefringence. The modes in crystal birefringence are determined by solving the characteristic equation $\Lambda_{ij} \varepsilon_j =0$, where $\varepsilon_i$ represents the mode polarization vector and the matrix $\Lambda_{ij}$ is given by
\begin{equation}
   \Lambda_{ij}= \left(
        \begin{array}{ccc}
            n_1^2 -n^2 \cos^2 \theta & 0 & n^2 \cos \theta \sin \theta \\
            0 & n_2^2 -n^2 & 0 \\
            n^2 \cos \theta \sin \theta & 0 & n_3^2 -n^2 \sin^2 \theta
        \end{array}
    \right), \label{cb1}
\end{equation}
with $n$ being the refractive index of the medium for the propagation of the probe light and $n_i$ being the principal refractive indices of the crystal (\cite{Fowles1989Introduction}). It is assumed that the coordinate axes are aligned with the principal axes of the crystal, and the probe light's propagation direction is $(\sin \theta,0,\cos \theta)$. Note that the principal indices are determined solely by the material properties, irrespective of the probe light's propagation direction. 

In fact, one can reproduce the characteristic matrix for the PM Lagrangian with a uniform magnetic field (\cite{Kim202231}) by making the following substitution in equation (\ref{cb1}):
\begin{equation}
    n_1^2=1,\quad n_2^2 = \frac{1-\eta_1 B^2}{1 - \eta_1 B^2 (1+2 \sin^2 \theta)}, \quad n_3^2 = \frac{1 + (2\eta_2 - \eta_1) B^2}{1 - \eta_1 B^2}, \label{cb2}
\end{equation}
where the magnetic field is directed along the $z$-axis. Now, the second \textit{effective} principal index, $n_2$ depends on the probe light's propagation direction, which is a notable difference from the case of crystal birefringence. This implies that the response of virtual electron-positron pairs in the vacuum can be more involved than that of bound electrons in crystals. As the PM Lagrangian is the generic form of non-linear Lagrangian in the weak field limit, such complication can be regarded as a generic feature of the field-modified vacuum.

\section{Conclusions and discussion}

\label{conclusions}

We have investigated the quantum refraction effects on the propagation and polarization of a photon 
in the dipole magnetic field background in pulsar emission, based on the PM Lagrangian from the HES one-loop action. 
Our main results are given by (\ref{nx})-(\ref{ny}) and (\ref{X})-(\ref{Y}) (for the effects on the propagation) 
and by (\ref{ex})-(\ref{ey}) and (\ref{ex2})-(\ref{ey2}) (for the effects on the polarization); via perturbation analysis, 
we have determined the leading-order corrections to both the propagation and polarization vectors due to quantum refraction, which result 
in the deflection of the propagation (as shown by Fig. \ref{fig2}) and the change of the two polarization modes 
(as shown by Figs. \ref{fig3} and \ref{fig3.2}). Further, we have inspected how quantum refraction affects the orthogonality 
between the propagation and polarization vectors and the Faraday rotation angle, the results of which are given in the texts 
at the end of Sections \ref{casei} and \ref{caseii} for the two polarization modes. From the dual refractive index and 
the associated polarization modes under the effect of quantum refraction, we have discussed quantum birefringence, with the optical 
phenomenology analogous to classical birefringence.

It is interesting to compare our analysis with other similar studies on the same topic. For instance, among others,  
\citet{Heyl2000Polarization} have set up the evolution equation $\partial \mathbf{s}/{\partial r}=\mathbf{\hat{\Omega}}\times \mathbf{s%
}$ for the `Stokes vector' $\mathbf{s}$, where the `birefringent vector' $\mathbf{\hat{\Omega}}$ 
contains the information about the propagation of a photon (e.g., from pulsar radiation) through 
an inhomogeneous birefringent medium (e.g., a strongly magnetized vacuum in the pulsar magnetosphere). 
Solving this equation, they have determined the vacuum QED effect from the strong magnetic field on the decoupling of
the polarization modes in pulsar emission. Their analysis can be applied to a pulsar environment wherein emission of radiation 
can take place at various energy scales; the QED effect may enable observations to distinguish between different mechanisms of pulsar emission and to reconstruct the structure of the magnetosphere. 
In contrast, we have shown the QED effect on the photon polarization in a different way, by solving equation (\ref{pol4}) for the transport of the polarization vector 
along the photon propagation. Despite the difference between the two approaches, the QED effects on the photon polarization might be amenable to the same interpretation, 
that is, the Faraday rotation (\citet{Caiazzo2019QED}). It will be of great interest to extend our study to a more general situation, where we cover various wavebands of radiation from an oblique rotator and investigate the frequency-dependent effect on the polarization modes. This is considered for our follow-up studies. 

It should be noted that not only the quantum refraction due to strong fields but also the plasmas spreading around the magnetosphere might affect the propagation and polarization of radiation. In a broad spectrum of curvature radiation, the low-frequency part, e.g., radio emission, is dominated by the plasma effects, while the high-frequency part, e.g., X-ray emission, is dominated by the quantum refraction effects, as is well discussed in \cite{Wang2007Wave}. To take the plasma effects into account properly, a good plasma distribution model is required, which is hardly available except for specific computational simulations (\cite{Petri2016Theory}). In contrast, our analysis for the quantum refraction effects is well-defined given a specific configuration of coherent magnetic fields (\cite{Kim2021General}). Therefore, for observation of quantum refraction effects, X-ray emission is more favoured than other emissions in the lower frequency regime; for example, a space telescope (the Imaging X-ray Polarimetry Explorer) performed the X-ray  polarimetry ($\sim$ a few $\mathrm{keV}$) to confirm the quantum refraction effects for a magnetar (\cite{Taverna2022Polarized}).

In this analysis we have not considered the effects of gravitation around the neutron star. As implied by Fig. \ref{fig1}, the magnetic field geometry would be affected by gravitation 
due to the neutron star mass; the closer to the neutron star surface, the stronger the gravitational effect is, which results in the field lines being more curved inwards. Consequently, the curvature radiation 
produced along the field lines would also be affected; its pulse profiles would change as the intensity of the radiation field would increase due to the effect of gravitation (\citet{Kim2021General}). 
In addition, the trajectories of photons would be affected by gravitation too; they should follow the `geodesics' of a curved spacetime geometry. For example, \citet{Heyl2003highenergy} and \citet{Caiazzo2019QED} 
have taken this into consideration in studying the evolution of the photon polarization under the vacuum QED effect in pulsar emissions. Inclusion of these gravitational effects 
is also considered for our follow-up studies.

\section*{Acknowledgements}

D.-H.K was supported by the Basic Science Research Program through the National Research Foundation
of Korea (NRF) funded by the Ministry of Education (NRF-2021R1I1A1A01054781). C.M.K. and S.P.K. were 
supported by Institute for Basic Science (IBS) under IBS-R012-D1. C.M.K. was also supported by Ultrashort Quantum
Beam Facility operation program (140011) through APRI, GIST and GIST Research Institute (GRI) grant funded by GIST. 
S.P.K. was also in part supported by National Research Foundation of Korea (NRF) funded by the Ministry of Education (NRF-2019R1I1A3A01063183).


\section*{Data Availability}

The inclusion of a Data Availability Statement is a requirement for articles
published in MNRAS. Data Availability Statements provide a standardised
format for readers to understand the availability of data underlying the
research results described in the article. The statement may refer to
original data generated in the course of the study or to third-party data
analysed in the article. The statement should describe and provide means of
access, where possible, by linking to the data or providing the required
accession numbers for the relevant databases or DOIs.



\bibliographystyle{mnras}
\bibliography{refs_VB_master}




\appendix


If you want to present additional material which would interrupt the flow of
the main paper, it can be placed in an Appendix which appears after the list
of references.


\section{Kinematic properties of a twisted curve and the Frenet--Serret
formulas}

\label{app}

At the end of Section \ref{dlr}, the trajectory of our light ray has been
identified as a twisted curve. Formally, kinematic properties of a twisted
curve can be interpreted by means of the Frenet--Serret formulas (%
\citet{Spivak1999Comprehensive}), which are known to be%
\begin{equation}
\frac{\mathrm{d}}{\mathrm{d}s}\left[ 
\begin{array}{c}
\mathbf{T} \\ 
\mathbf{N} \\ 
\mathbf{B}%
\end{array}%
\right] =\left[ 
\begin{array}{ccc}
0 & \kappa & 0 \\ 
-\kappa & 0 & \tau \\ 
0 & -\tau & 0%
\end{array}%
\right] \left[ 
\begin{array}{c}
\mathbf{T} \\ 
\mathbf{N} \\ 
\mathbf{B}%
\end{array}%
\right] ,  \label{A1}
\end{equation}%
where $\mathbf{T}$, $\mathbf{N}$ and $\mathbf{B}$ are named the unit
tangent, normal and bi-normal vectors, respectively, which collectively
constitute the Frenet--Serret frame that forms an orthonormal basis of
3-space, and $s$ is an affine parameter to measure the arc length along the
curve, and $\kappa $ is the curvature and $\tau $ is the torsion of the
curve. Intuitively, $\kappa $ measures the failure of the curve to be a
straight line, while $\tau $ measures the failure of the curve to be planar;
in regard to our case, the former refers to the trajectory being deflected
from a straight line as viewed in the $xz$-plane (due to the quantum
refraction effect), while the latter refers to the trajectory following a
parabolic path in another plane perpendicular to the $xz$-plane, hence
failing to stay in a single plane (due to the rotational effect of the
pulsar magnetosphere).

The Frenet--Serret analysis is available to a curve which is expressed by a
well-behaved $C^{k}$ function with $k\geq 3$ everywhere it is defined.
However, the trajectory curve of our light ray, $\mathbf{r}=\left(
X,Y,Z\right) $, as given by equations (\ref{X})-(\ref{Y}), is not well defined in
this sense, and therefore we cannot resort to the Frenet--Serret formulas to
construct the orthonormal basis $\left\{ \mathbf{T},\mathbf{N},\mathbf{B}%
\right\} $ out of the curve. Instead, we use rather a geometrically
intuitive approach: (i) First, build $\mathbf{T}$ from $\mathbf{\hat{n}/}%
\left\vert \mathbf{\hat{n}}\right\vert $ by its definition, (ii) Then, build 
$\mathbf{N}$ by taking $N_{y}$ the same as $T_{y}$ and by determining the $%
N_{x}$ and $N_{z}$ from $\mathbf{N\cdot T}=0$ and $\left\vert \mathbf{N}%
\right\vert =1$ (so that the curvature of the curve be defined from the
deflection in the $xz$-plane), (iii) Lastly, build $\mathbf{B}$ from $%
\mathbf{T}\times \mathbf{N}$. After some tedious calculations, they are
obtained as 

\begin{align}
\mathbf{T}& \equiv \frac{\mathbf{\hat{n}}}{\left\vert \mathbf{\hat{n}}%
\right\vert }  \notag \\
& \approx \left( \hat{n}_{x\left[ 0\right] }+\frac{\hat{n}_{z\left[ 0\right]
}^{2}-\hat{n}_{x\left[ 0\right] }^{2}}{\hat{n}_{x\left[ 0\right] }}\delta n_{%
\left[ 1\right] }\right) \mathbf{e}_{x}+\hat{n}_{y\left[ 0\right] }\left(
1-2\delta n_{\left[ 1\right] }\right) \mathbf{e}_{y}+\left( \hat{n}_{z\left[
0\right] }+\frac{\hat{n}_{x\left[ 0\right] }^{2}-\hat{n}_{z\left[ 0\right]
}^{2}}{\hat{n}_{z\left[ 0\right] }}\delta n_{\left[ 1\right] }\right) 
\mathbf{e}_{z}+\mathcal{O}\left( \left( \Omega r_{\mathrm{o}}/c\right)
^{2},_{\left[ 2\right] }\right) ,  \label{A2}
\end{align}%
%
%
%
%
%
%
%
%
%
%
%
%
%
%
%
%
%
%
%
%
%
%

\begin{equation}
\mathbf{N}\approx \left( \hat{n}_{z\left[ 0\right] }+\frac{\hat{n}_{x\left[ 0%
\right] }^{2}-\hat{n}_{z\left[ 0\right] }^{2}}{\hat{n}_{z\left[ 0\right] }}%
\delta n_{\left[ 1\right] }\right) \mathbf{e}_{x}+\hat{n}_{y\left[ 0\right]
}\left( 1-2\delta n_{\left[ 1\right] }\right) \mathbf{e}_{y}-\left( \hat{n}%
_{x\left[ 0\right] }+\frac{\hat{n}_{z\left[ 0\right] }^{2}-\hat{n}_{x\left[ 0%
\right] }^{2}}{\hat{n}_{x\left[ 0\right] }}\delta n_{\left[ 1\right]
}\right) \mathbf{e}_{z}+\mathcal{O}\left( \left( \Omega r_{\mathrm{o}%
}/c\right) ^{2},_{\left[ 2\right] }\right) ,  \label{A3}
\end{equation}%
%
%
%
%
%
%
%
%
%
%
%
%
%
%
%
%
%
%
%
%
%
%

\begin{align}
\mathbf{B}&\equiv \mathbf{T}\times \mathbf{N}  \notag \\
&\approx \hat{n}_{y\left[ 0\right] }\left[ -\hat{n}_{x\left[ 0\right] }-%
\hat{n}_{z\left[ 0\right] }+\left( 4\hat{n}_{x\left[ 0\right] }+4\hat{n}_{z%
\left[ 0\right] }-\frac{1}{\hat{n}_{x\left[ 0\right] }}-\frac{1}{\hat{n}_{z%
\left[ 0\right] }}\right) \delta n_{\left[ 1\right] }\right] \mathbf{e}_{x}+%
\mathbf{e}_{y}  \notag \\
& \hspace{10pt}+\hat{n}_{y\left[ 0\right] }\left[ \hat{n}_{x\left[ 0\right] }-\hat{n}_{z%
\left[ 0\right] }+\left( -4\hat{n}_{x\left[ 0\right] }+4\hat{n}_{z\left[ 0%
\right] }+\frac{1}{\hat{n}_{x\left[ 0\right] }}-\frac{1}{\hat{n}_{z\left[ 0%
\right] }}\right) \delta n_{\left[ 1\right] }\right] \mathbf{e}_{z}+\mathcal{%
O}\left( \left( \Omega r_{\mathrm{o}}/c\right) ^{2},_{\left[ 2\right]
}\right) ,  \label{A4}
\end{align}%
%
%
%
%
%
%
%
%
%
%
%
%
%
%
%
%
%
%
%
%
%
%
where

\begin{equation}
\delta n_{\left[ 1\right] }\approx \frac{4\eta _{2}B_{\mathrm{o}}^{2}\left(
\cos \theta _{\mathrm{o}}\hat{n}_{x\left[ 0\right] }-\sin \theta _{\mathrm{o}%
}\hat{n}_{z\left[ 0\right] }\right) ^{2}r_{\mathrm{o}}^{6}s^{2}}{\left(
3\cos ^{2}\left( \theta _{\mathrm{o}}-\alpha \right) +1\right) ^{2}\left[
s^{2}+2\left( \cos \theta _{\mathrm{o}}\hat{n}_{z\left[ 0\right] }+\sin
\theta _{\mathrm{o}}\hat{n}_{x\left[ 0\right] }\right) r_{\mathrm{o}}s+r_{%
\mathrm{o}}^{2}\right] ^{4}},  \label{A5}
\end{equation}%
%
%
%
%
%
%
%
%
%
%
%
%
%
%
%
%
%
%
%
%
%
%
taken from equation (\ref{dn1}) with equations (\ref{xs}), (\ref{zs}) and $B_{\mathrm{o%
}}=\mu \left( 3\cos ^{2}\left( \theta _{\mathrm{o}}-\alpha \right) +1\right)
^{1/2}/r_{\mathrm{o}}^{3}$ substituted, and $\hat{n}_{x\left[ 0\right] }$, $%
\hat{n}_{z\left[ 0\right] }$ and $\hat{n}_{y\left[ 0\right] }$\ are given by
equations (\ref{nox})-(\ref{noy}), respectively. One can easily check out $%
\left\vert \mathbf{T}\right\vert \approx \left\vert \mathbf{N}\right\vert
\approx \left\vert \mathbf{B}\right\vert \approx 1+\mathcal{O}\left( \left(
\Omega r_{\mathrm{o}}/c\right) ^{2},_{\left[ 2\right] }\right) $ and $%
\left\vert \mathbf{T\cdot N}\right\vert \approx \left\vert \mathbf{N\cdot B}%
\right\vert \approx \left\vert \mathbf{B\cdot T}\right\vert \approx 0+%
\mathcal{O}\left( \left( \Omega r_{\mathrm{o}}/c\right) ^{2},_{\left[ 2%
\right] }\right) $, as desired.

\begin{figure*}
\subfloat[]{
		\includegraphics[angle=0, width=9.4653cm]{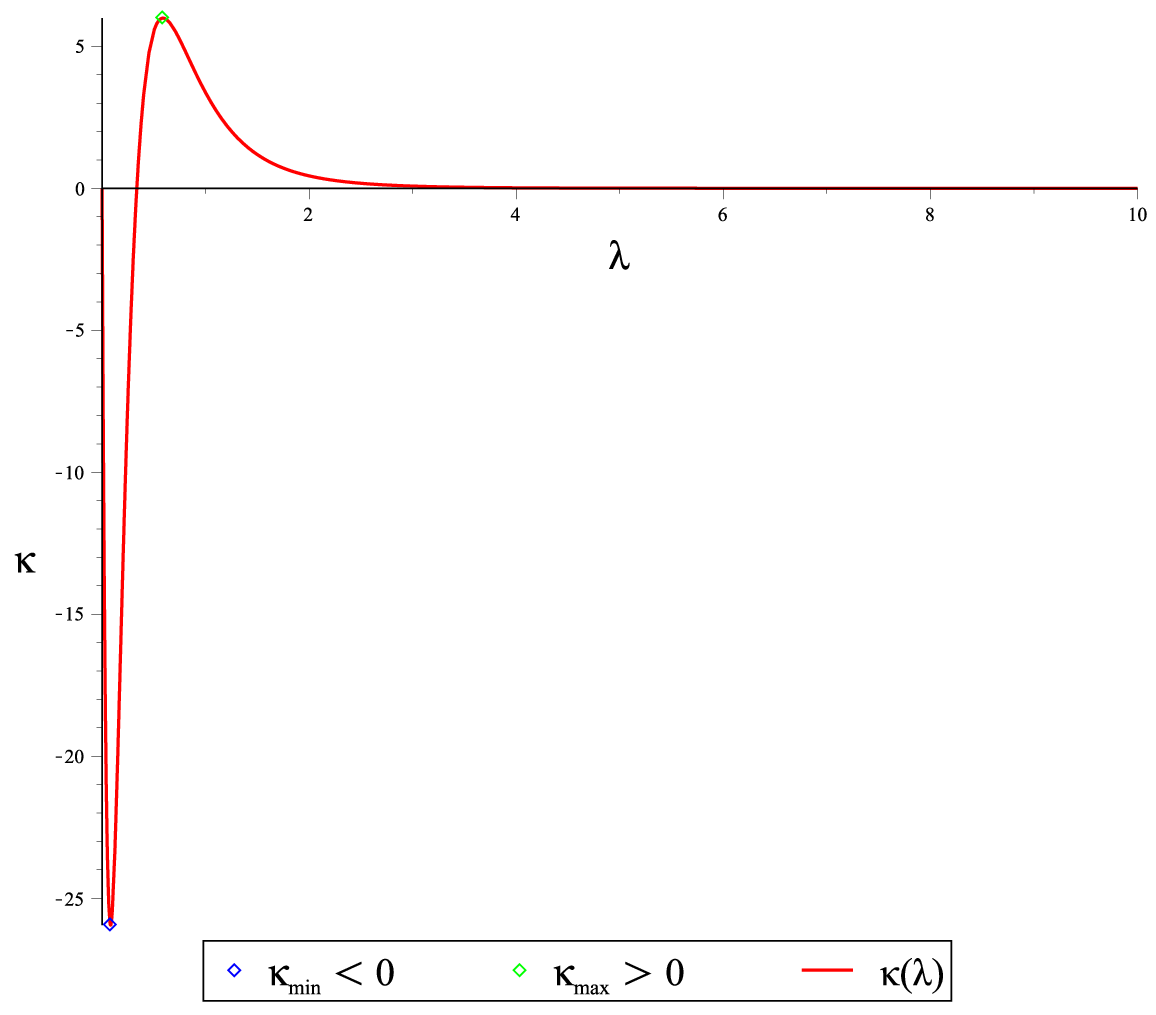}
		\label{fig:subfig1}} 
\subfloat[]{
		\includegraphics[angle=0,width=8.8766cm]{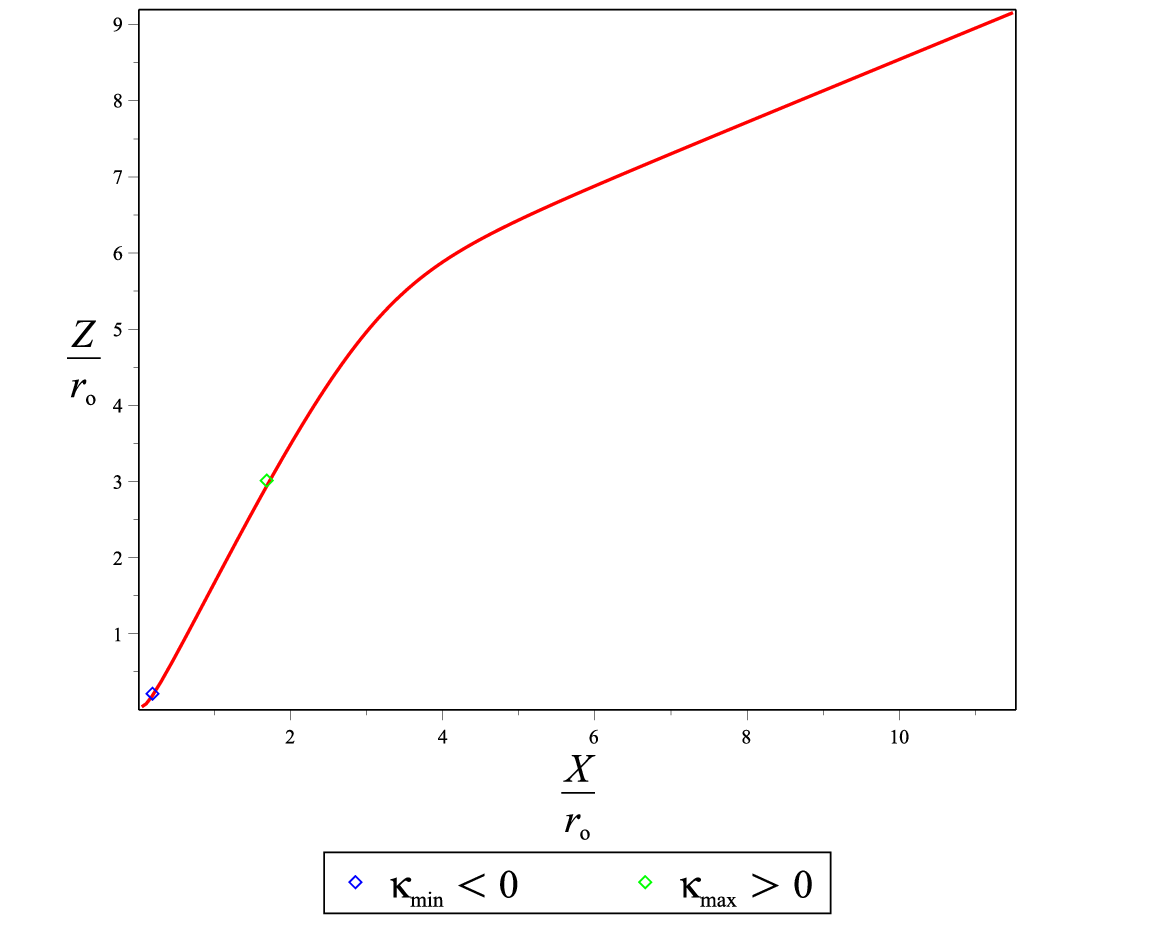}
		\label{fig:subfig2}} 
\caption{(a) The curvature $\protect\kappa $ (dimensionless) plotted against 
$0\leq \protect\lambda \leq 10$, (b) The curve $\left( X/r_{\mathrm{o}
},0,Z/r_{\mathrm{o}}\right) $ in the $xz$-plane, corresponding to the
curvature $\protect\kappa \left( \protect\lambda \right) $.}
\label{fig4}
\end{figure*}

Following \citet{Born1999Principles}, we can determine the curvature $\kappa 
$ of our curve $\mathbf{r}=\left( X,Y,Z\right) $ as given by equations~(\ref{X})-(%
\ref{Y}): 
\begin{equation}
\kappa =\frac{1}{\rho }=\mathbf{N\cdot }\nabla \left( \ln n\right) ,
\label{A8}
\end{equation}%
where $\rho $ denotes the radius of curvature and $\mathbf{N}$ refers to the
unit normal vector given by equation~(\ref{A3}) above, and the refractive index $n
$ is given by equation~(\ref{n2}). Similarly as in Section~\ref{cplr}, we can work
out from equation~(\ref{A8}), 
\begin{align}
\kappa &=\mathbf{N}_{\left[ 0\right] }\cdot \nabla \left( \delta n_{\left[ 1%
\right] }\right)   \notag \\
& \approx \frac{8\eta _{2}B_{\mathrm{o}}^{2}\left( \cos \theta _{\mathrm{o}}%
\hat{n}_{x\left[ 0\right] }-\sin \theta _{\mathrm{o}}\hat{n}_{z\left[ 0%
\right] }\right) ^{2}\left( \hat{n}_{x\left[ 0\right] }^{2}-\hat{n}_{z\left[
0\right] }^{2}\right) r_{\mathrm{o}}^{6}}{\left( 3\cos ^{2}\left( \theta _{%
\mathrm{o}}-\alpha \right) +1\right) ^{2}\hat{n}_{x\left[ 0\right] }\hat{n}%
_{z\left[ 0\right] }}\frac{\left[ 3s^{2}+2\left( \cos \theta _{\mathrm{o}}%
\hat{n}_{z\left[ 0\right] }+\sin \theta _{\mathrm{o}}\hat{n}_{x\left[ 0%
\right] }\right) r_{\mathrm{o}}s-r_{\mathrm{o}}^{2}\right] s}{\left[
s^{2}+2\left( \cos \theta _{\mathrm{o}}\hat{n}_{z\left[ 0\right] }+\sin
\theta _{\mathrm{o}}\hat{n}_{x\left[ 0\right] }\right) r_{\mathrm{o}}s+r_{%
\mathrm{o}}^{2}\right] ^{5}},  \label{A9}
\end{align}%
where $\mathbf{N}_{\left[ 0\right] }$ denotes the unperturbed (classical)
part of $\mathbf{N}$, taken from equation (\ref{A3}) with all the terms of $%
\delta n_{\left[ 1\right] }$ removed, and $\delta n_{\left[ 1\right] }$, $%
\hat{n}_{x\left[ 0\right] }$ and $\hat{n}_{z\left[ 0\right] }$ are given by
equations (\ref{A5}), (\ref{nox}) and (\ref{noz}), respectively. Here we see that
the curvature $\kappa $ is solely due to the quantum refraction effect,
being led by the parameter $\eta _{2}B_{\mathrm{o}}^{2}$. In Fig. \ref%
{fig:subfig1} is plotted the curvature $\kappa $ (dimensionless, multiplied
by $r_{\mathrm{o}}$) against $0\leq \lambda \leq 10$ (with the substitution $%
s=r_{\mathrm{o}}\lambda $ in equation (\ref{A9})), and in Fig. \ref{fig:subfig2}
is plotted the corresponding curve $\left( X/r_{\mathrm{o}},0,Z/r_{\mathrm{o}%
}\right) $, which is projected onto to the $xz$-plane. Here we assume $r_{%
\mathrm{o}}=2\times 10^{6}\,\mathrm{cm}$, $\theta _{\mathrm{o}}=60^{\circ }$%
, $\alpha =45^{\circ }$, $\Omega =2\pi \times 10^{2}\,\mathrm{Hz}$ and $\eta
_{2}B_{\mathrm{o}}^{2}\sim 10^{4}$ ($10^{9}$ times as
large as an actual order $\sim 10^{-5}$; fairly exaggerated for intuitive
visualization). Note the two points marked for $\kappa _{\min }$ and $\kappa
_{\max }$ in each plot; in particular, the trajectory is curved the most
downward and the most upward at the former and the latter points, respectively.

\bsp
\label{lastpage}
\end{document}